\documentclass[dvipsnames]{aastex63}

\shorttitle{Probing QSO and DLA halo mass}
\shortauthors{X.Lin,Z.Cai.,et al.}
\graphicspath{{./}{figures/}}
\usepackage{verbatim}
\usepackage{amsmath}
\usepackage{appendix}
\usepackage{multirow}
\usepackage{threeparttable}
\usepackage{enumitem}
\setlist[enumerate,itemize]{topsep=0.5ex,itemsep=0pt,parsep=0.5ex}
\SetLabelAlign{CenterWithParen}{\hfil(\makebox[1.0em]{#1})\hfil}
\usepackage{microtype}
\usepackage{ulem}
\usepackage{bm}
\usepackage{hyperref}

\newcommand{\fsky}{f_\mathrm{sky}}
\newcommand{\bqso}{b_\mathrm{QSO}}
\newcommand{\bdla}{b_\mathrm{DLA}}

\newcommand{\BOSSaqso}{0.71\pm0.19}
\newcommand{\BOSSbqso}{2.55\pm0.70}
\newcommand{\BOSSmqso}{11.79^{+0.63}_{-0.40}}

\newcommand{\DLAaqso}{0.68\pm0.18}
\newcommand{\DLAbdla}{1.37^{+1.30}_{-0.92}}
\newcommand{\DLAmdla}{10.60^{+1.44}_{-5.27}}
\newcommand{\upbdla}{3.1}
\newcommand{\upmdla}{12.3}

\begin{document}

\title{Constraining the Halo Mass of Damped Ly$\alpha$ Absorption Systems (DLAs) at $z=2-3.5$ using the Quasar-CMB Lensing Cross-correlation}

\author{Xiaojing Lin}
%\affiliation{Department of Astronomy, 
%Tsinghua University, Beijing 100084, China}
\affiliation{Department of Astronomy, School of Physics, Peking University, Beijing 100871, China}
\email{linxiaojing@pku.edu.cn}

\author{Zheng Cai}
\affiliation{Department of Astronomy, 
Tsinghua University, Beijing 100084, China}
\email{zcai@mail.tsinghua.edu.cn}

\author{Yin Li}
\affiliation{Center for Computational Astrophysics \&
Center for Computational Mathematics,
Flatiron Institute, 162 5th Avenue, 10010, New York, NY, USA}
\email{yinli@flatironinstitute.org}

\author{Alex Krolewski}
\affiliation{Department of Astronomy, University of California, Berkeley, CA 94720}
\affiliation{Physics Division, Lawrence Berkeley National Laboratory, Berkeley, CA}

\author{Simone Ferraro}
\affiliation{Physics Division, Lawrence Berkeley National Laboratory, Berkeley, CA}




\begin{abstract}
We study the cross correlation of damped Ly$\alpha$ systems (DLAs) and
their background quasars, using the most updated DLA catalog and the  \textit{Planck} 2018 CMB lensing
convergence field.
Our measurement suggests that the DLA bias $\bdla$ is smaller than
$\upbdla$, corresponding to $\log(M/M_\odot h^{-1})\leq\upmdla$ at a
confidence of 90\%.
These constraints are broadly consistent with \citet{JCAP_DLA} 
and previous 
measurements by cross-correlation between DLAs and the Ly$\alpha$ forest
\citep[e.g.][]{Lyman_forest_2012,Lyman_forest_2018}.
Further, our results demonstrate the potential of obtaining a
more precise measurement of the halo mass of high-redshift sources
using next generation CMB experiments with a higher angular resolution. The python-based codes and data products of our analysis are available at \href{https://github.com/LittleLin1999/CMB-lensingxDLA}{https://github.com/LittleLin1999/CMB-lensingxDLA}.
\end{abstract}



\section{Introduction} \label{sec:intro}

Damped Ly$\alpha$ systems (DLAs) are a class of absorbers along the QSO
sight lines with high neutral hydrogen (HI) column densities of
$N_{\rm{HI}}\geq2\times10^{20}$ cm$^{-2}$ \citep{Wolfe_1986}.
They are featured by their damping wings absorption profile, a result of
the quantum natural broadening of the HI Ly$\alpha$ transition.
DLAs are thought to dominate the neutral-gas content of the Universe in
the redshift range of $0 \leq z\leq 5$ \citep{Wolfe_2005}, acting as
important neutral gas reservoirs for star formation \citep[e.g.][]{Nagamine_2004,Ota_2014,Rudie_2017}.
%\yl{``high'' means $0\sim5$ or higher?}
They can also be a powerful cosmological probe for related research,
such as studying the hosts of quasars
\citep[e.g.][]{Hennawi_2009,Zafar_2011,Cai_2014}.

%The precise nature of DLAs remains unknown,
Over the last three decades, a number of attempts have been made to
discover the nature of DLAs by directly probing the emission of the DLA
galaxies using the largest optical or infrared telescopes
\citep[e.g.][]{Le_Brun_1997,Moller_2002,Chen_2005}.
These studies have been moderately successful at low redshift.
Nevertheless, after the many deep observations with innovative
techniques
\citep[e.g.][]{Kulkarni_2006,Fynbo2010,Fumagalli2015,Johnson2016}, only
about 20 DLA host galaxies at $z\gtrsim2$ have been identified.
These results suggest that DLA host galaxies have a relatively small,
sub-$L^*$ galaxies with the star formation rate (SFR) of $\sim 0.1 - 10\
M_\odot$ yr$^{-1}$ \cite[e.g.,][]{Krogager2017}.
With the Atacama Large Millimeter/submillimeter Array (ALMA),
%Neeeleman et al. (2018; 2020) provides a complementary approach
%, whereby we %can search for longer wavelength
%emission (e.g., from CO, [C ii] lines, and radio continuum), which is less affected by the presence of dust and
%arises predominantly from the molecular and atomic gas
%inside the galaxy
%With the progress of the observational efforts,
\citet{Neeleman2018,Neeleman2020} have revealed
a small sample ($<10$) of host galaxies of
high-metallicity DLAs at $z\approx4$ using the sub-millimeter
observations. They find that
the SFR of DLA hosts is on the order of 10 - 100 M$_\odot$ yr$^{-1}$, indicating
that metal-rich DLA hosts could be super-$L^*$ galaxies,
at odds with the previous findings.
%Nevertheless, the statistical host galaxy properties of DLAs remains unclear. On the theoretical side,
%There are still lacking of sufficient observational evidences for
%the mass distribution of their host halo masses and
%the type of their host galaxies.

Statistical studies of the DLA hosts also remain
unclear and controversial.
%The studying the clustering properties of DLAs is expected to
%provide strong constraints on their mass distribution. There
%are
Several works measure the DLA bias by
studying the cross-correlation between DLAs and other tracers,
such as Lyman break galaxies \citep[e.g.][]{Cooke_2006,Lee_2011} and Ly$\alpha$ forests \citep[e.g.][]{Lyman_forest_2012,Lyman_forest_2018}. The latter imply that DLAs are hosted by massive halos with typical masses of $10^{11}-10^{12}M_\odot$.
Mass-metallicity statistics reveal that the DLA hosts should have a stellar mass of $10^{8.5}\ M_\odot$ \citep[e.g.][]{Moller_2013}. This
result is also supported by several hydrodynamical simulations \citep[e.g.][]{DLA_sim_1,Berry2016} which suggest that DLA galaxies have stellar masses ranging from
$10^6\ M_\odot$ to $10^{11}\ M_\odot$, with the median value around
$10^8\ M_\odot$. %. These bias values are also higher than some model predictions \citep[e.g.][]{Mo_1998} and those inferred
More statistical efforts are
needed to provide more independent measurements on the DLA hosts.

Besides the detection techniques mentioned above, weak lensing of the
cosmic microwave background (CMB) is a powerful tool to probe halo mass
at high redshift.
%As CMB photons travel to the observer from the last scattering surface,
%they are gravitationally deflected by the matter, leading to the
%so-called weak lensing effect. Therefore, weak lensing of CMB is a good
%tracer of the dark matter field as well as large scale structures.
The cross-correlation between the CMB and other tracers of large-scale
structure
%covering narrow redshift ranges
can be used to measure the tracer bias
%\yl{\sout{cosmic } bias of the tracer distribution with respect to that of the dark matter}
, with which one
can infer the typical halo masses of the tracer populations.
The cross-correlation measurement is not prone to systematic effects
that are possibly present in the auto-correlation approach, such as
incompleteness, random catalog generation, masking,
etc.\citep{Reid_2015,Geach_2019}.
%{(repeated in Sec Discussion)}
This technique has been performed to analyse the clustering properties
of quasars and galaxies and provide strong cosmological constraints
\citep[e.g.][]{Sherwin_2012,Bianchini_2015,Simone_mag_bias}.
Likewise, CMB photons are deflected not only by the quasar hosts, but
also by DLAs in the sightlines.
The measured gravitational magnification of the background quasars would
be strongly affected by the foreground DLAs if DLAs reside in massive
halos comparable to those of the quasar hosts.
The assumption has been confirmed by studies on the CMB lensing-quasar
cross-correlations using quasars with and without DLAs in their
sightlines \citep{JCAP_DLA}.

In this paper we carefully measure the cross-correlation between the \textit{Planck} CMB lensing convergence map and two quasar overdensity maps. One of the two quasar maps ensures that each quasar sight line contains at least one DLA. Through comparing the cross-correlation results of the two maps, the properties of DLA host are extracted. In \S~\ref{sec:data} we introduce the data samples used in this work. \S~\ref{sec:method} presents the estimator of the cross-correlation power spectrum and its error. In \S~\ref{sec:results} we measure the biases of the quasar and DLA samples. The null test results are reported in \S~\ref{sec:null}. In \S~\ref{sec:discuss} we discuss our measurements and their consistency with several previous works. We draw our conclusions in \S~\ref{sec:conclusion}. Additionally, we apply simple tests by applying this approach to a quasar bias evolution model, a tSZ-free CMB lensing map and another DLA catalog in Appendix \ref{Laurent17}, \ref{non-tSZ}, and \ref{Mingfeng}.
Throughout the paper, we adopt a flat $\Lambda$ cold dark matter ($\Lambda$CDM) cosmological model as described in \cite{planck18_cosmo}, with $H_0$=67.4, $\Omega_\mathrm{m}$=0.315, $\Omega_\mathrm{b}h^2$=0.0224, $\Omega_\mathrm{c}h^2=0.120$.

\section{DATA SAMPLE}
\label{sec:data}

\subsection{BOSS quasars}

Both of our QSOs and DLAs are selected from the third stage of
the SDSS survey \citep[SDSS III,][]{SDSS-III}.
SDSS-III mapped a total of 14,555 unique square degrees of the sky,
including contiguous areas of $\sim7,500 ~\rm{deg}^2$ in the North
Galactic Cap (NGC) and $\sim3,100 ~\rm{deg}^2$ in the South Galactic Cap
(SGC) using the Sloan Foundation 2.5-meter Telescope at Apache Point
Observatory in New Mexico.
As the main dark time survey of SDSS III, the Baryon Oscillation
Spectroscopic Survey (BOSS) \citep[BOSS,][]{BOSS_survey} observed about 300,000 quasars, and 184,000 of which fall
within the redshift range of $2.15 \leq z \leq 4$.
BOSS detected the
baryon acoustic oscillation (BAO) using HI absorption lines in the
intergalactic medium (IGM) at z$\sim$2.5 \citep{Busca_2013}.
%\yl{is this what we want to emphasize?}.
%{\bf (No, just a brief introduction to BOSS) }
%\yl{It sounds like this is the only or most important thing BOSS does.
%But there are many more. I would at least swap this and the next
%sentences.}.
%and the spatial distribution of luminous red galaxies at
%z$\sim$0.7 \citep{Anderson_2012,Alam_2017}.
%As part of the efforts to attain this goal, survey
%was designed to discover at least 15 quasars
%per square degree over the range of $2.15 \leq z \leq 3.5$ in order to
%map the large-scale structure traced by the Ly-$\alpha$ forest,
%and finally
%BOSS survey
%pushes much fainter than SDSS-I/II, reaching $g < 22.0$, or $r < 21.85$.
The quasar samples used in our work are from the final SDSS-III/BOSS
quasar catalog\footnote{\url{https://www.sdss.org/dr12/algorithms/boss-dr12-quasar-catalog/}} \citep{BOSS_DR12},
containing 184,101 quasars at $z \geq 2.15$, with 167,742 of them
being new discoveries.

\subsection{Planck 2018 data}

The Planck satellite, launched on 14 May 2009, mapped the
anisotropies of the cosmic microwave background (CMB) at multiple
frequencies with a high sensitivity and small angular resolution.
\textit{Planck} 2018 results \citep{Planck} provide variations of
lensing potential estimates as tables of spherical harmonic
coefficients up to $\ell = 4096$, as well as a lens reconstruction
analysis mask in the HEALPix format \citep{healpix} with $N_{\rm{side}}=2048$,
i.e.\ a pixel side of $d_{\mathrm{pix}} \approx 1.7$ arcmin. The lensing
convergence map\footnote{Lensing products are are available from
the Planck Legacy Archive: \url{https://pla.esac.esa.int/\#cosmology}}
we use was reconstructed by the minimum-variance (MV) estimate
\citep{planck_MV} from temperature and polarization, covering
approximately 70\% of the sky.

\subsection{DLA catalog}

We utilize a DLA catalog classified by a convolutional neural network
(CNN) based DLA finder \citep{DLA_CNN}.
The CNN architecture used multi-task learning to characterize strong HI
Ly$\alpha$ absorption in quasar spectra, and estimated the corresponding
redshift $z_{abs}$ and HI column density $N_{\rm{HI}}$.
% The accuracy of the $z_{\rm{abs}}$ and $N_{\rm{HI}}$ are xxx and xxx, respectively.
By comparing the absorption redshifts and $N_{\rm{HI}}$ values between
the matched DLAs from their algorithm and that in \citet{Garnett2017},
they measured small mean offsets of $\Delta z=-0.00035$ and $\Delta
N_{\rm{HI}}=-0.038$ and standard deviations of 0.002 and 0.16 dex
respectively.
The model also outputs a \textit{confidence} parameter to depict the
confidence level based on how precisely the CNN model predicted the
location of the DLA.
%\yl{precisely (low variance) or accurately (low bias)?}
%{\bf (Hmm...sorry I am not clear but I guess it probably indicates
%precision...In Parks et al. 2018 they said "The full scan provides a
%set of offset values which we histogram, expecting to observe a cluster
%of values at the center of a bonafide DLA... We define a confidence
%parameter where we sum the histogram over the nearest 5 pixels (i.e.
%$\pm$2 pixels) and take a 9-pixel median filter of that result limiting
%its maximum value to 1.")}
The model classified and measured absorption lines of DLAs in sightlines
in the SDSS-III/DR12 QSO database \citep{BOSS_DR12}.
The resulted catalog has 50,969 systems with $\log N_{\rm{HI}} \geq
20.3$, $z_{\rm{abs}} > 2$, and $z_{\rm{abs}} < z_{\rm{em}}$, in the
174,691 sightlines not flagged as broad absorption lines (BALs) in the
BOSS quasar catalog.

{ As a test we also apply our analysis to the DLA catalog released by \citet{Mingfeng_2020}. More details are present in Appendix \ref{Mingfeng}. }

\subsection{Sample selection}\label{sample_selection}

We apply a series of constraints on the BOSS catalog similar to that in
\cite{BOSS_auto}.
%, because we {\bf ?? I think we re-measured QSO bias using CMB lensing adopt their measured quasar bias as our fiducial bias model.} %limited their clustering measurements
%to the NGC region over the range of $2.2<z<3.4$.
One of our selection criteria is to reject quasars that are detected at
1.4 GHz by the Faint Images of the Radio Sky at Twenty-Centimeters
(FIRST) survey \citep{FIRST}.
Radio-loud quasars are removed because they may cluster differently from optically selected quasars\citep{Radio-loud1,Radio-loud2}.
Our selection criteria are the follows:
\begin{itemize}
  \item \textit{1st cut}: we include only quasars in the redshift
    range of $2.2<z<3.4$.
  \item \textit{2nd cut}: a magnitude cut of $m_{\rm g}<22$ is applied.
  \item \textit{3rd cut}: a luminosity cut of
    $-28.74<M_{\rm{i}}<-23.78$ is applied. $M_{\rm{i}}$ refers to the
    absolute i-band magnitude K-corrected to $z = 2$, and has been
    calculated for each quasar by SDSS\citep{Richards2006}.
  \item \textit{4th cut}: we limit our measurements to the NGC
    region, because some previous analyses
    \citep[e.g.][]{White_2012,BOSS_auto} have reported unexplained
    differences between clustering measurements in the BOSS NGC and SGC
    regions.
  \item \textit{5th cut}: quasars detected by the FIRST survey are
    excluded.
\end{itemize}
For DLA samples, we additionally eliminate broad absorption line (BAL)
systems, whose profiles can be easily confused with the Voigt profiles
of DLAs, and apply a confidence cut of \textit{confidence} $\geq 0.3$
\citep{DLA_CNN}, besides the constraints above.
The total sample size of the subset of BOSS quasars is 105,642.
And the final DLA catalog includes 17,774 quasars with 20,848 DLAs in
their sightlines.

\begin{figure*}[htbp]
\gridline{
        \fig{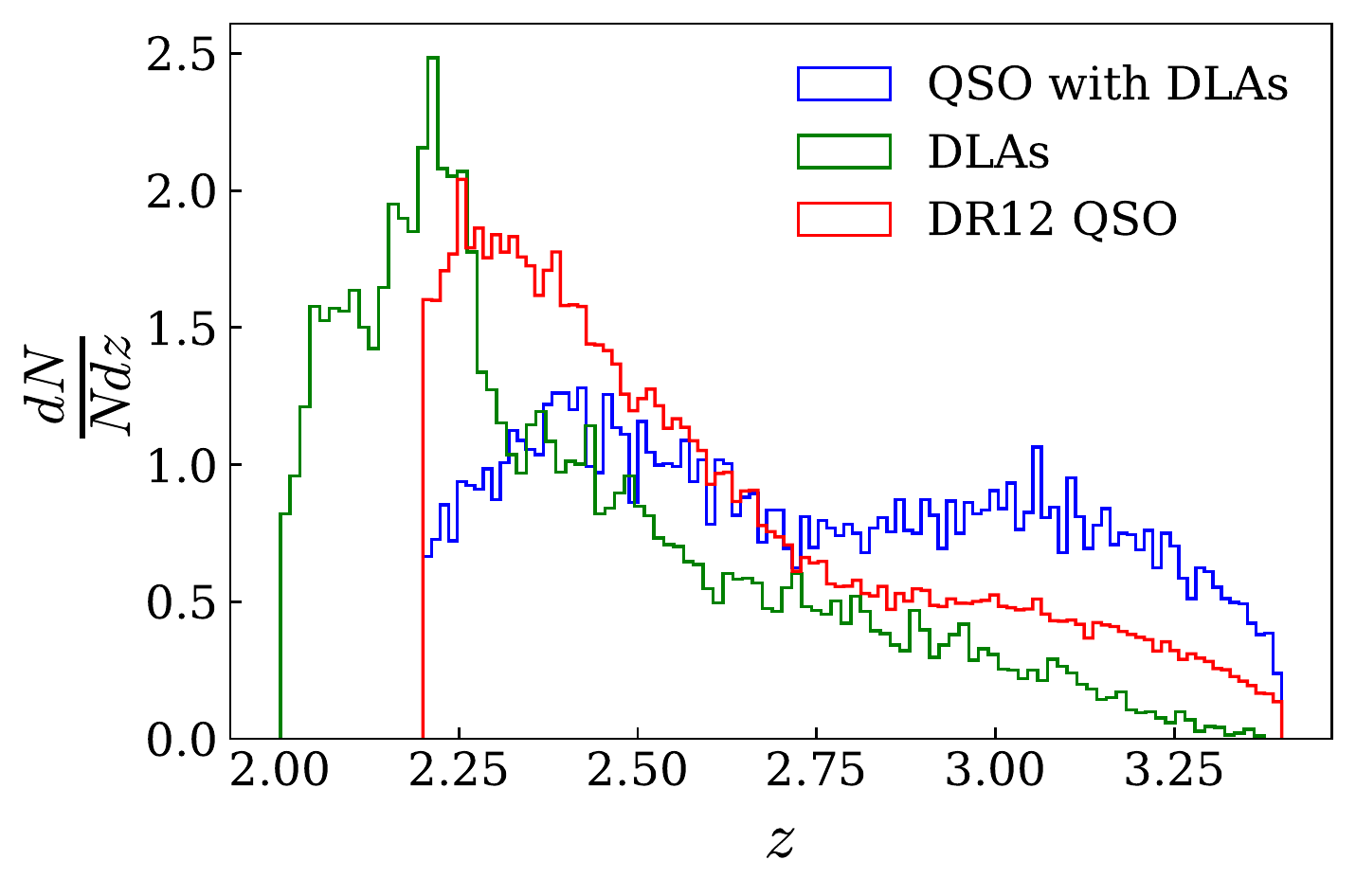}{0.33\textwidth}{Redshift distribution}
        \fig{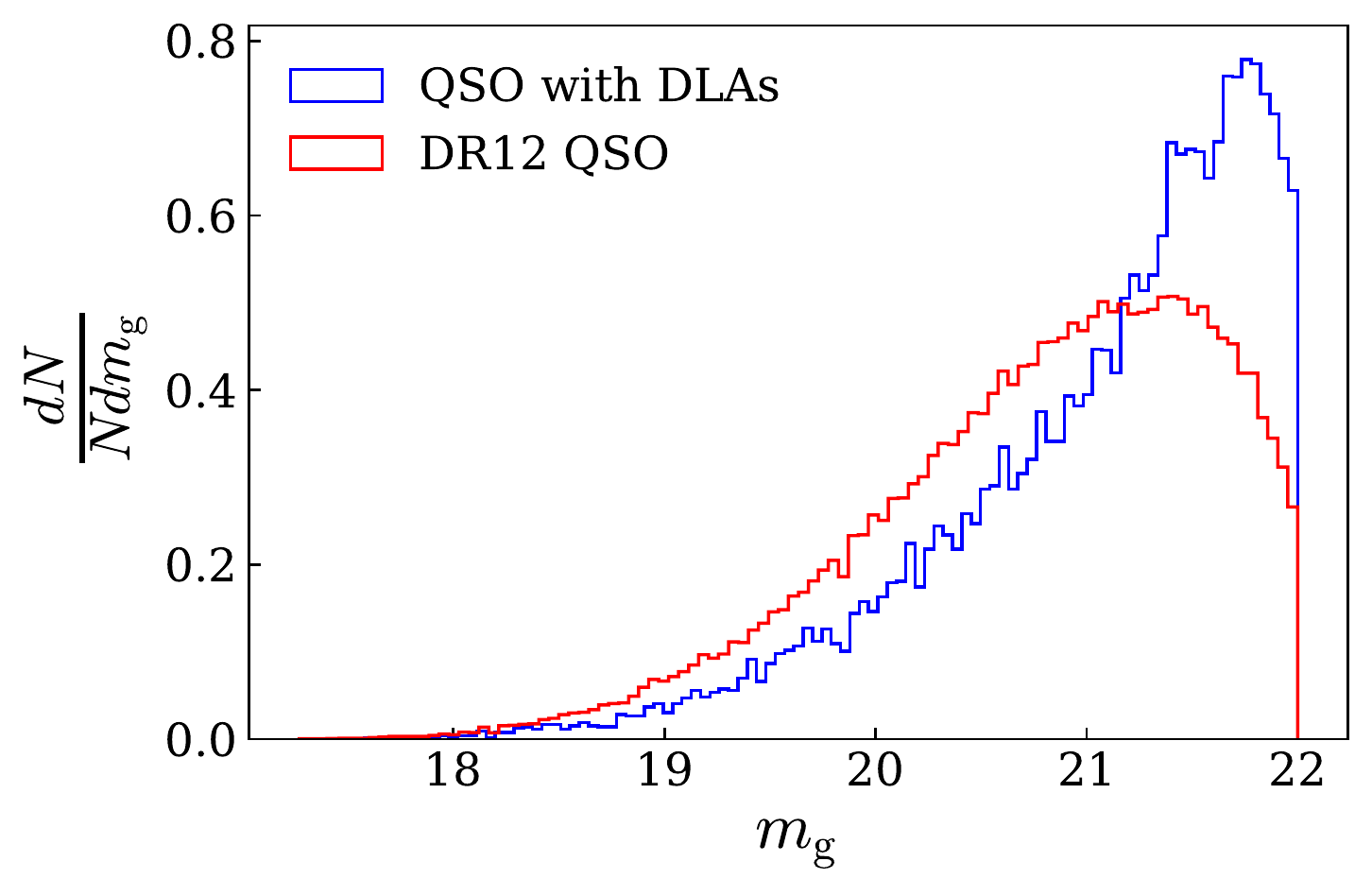}{0.33\textwidth}{Magnitude distribution}
        \fig{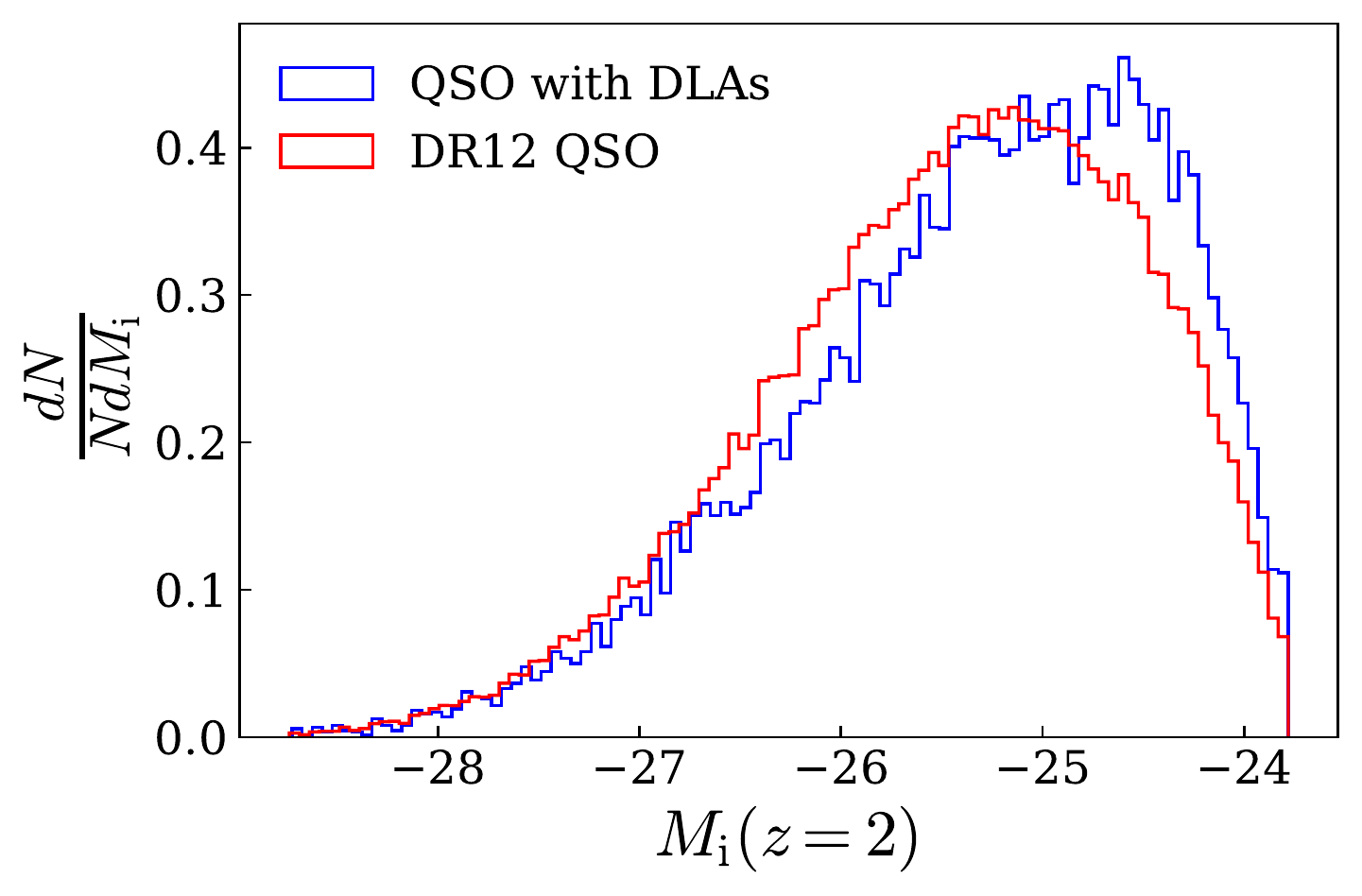}{0.33\textwidth}{Luminosity distribution}
          }
\caption{The normalised redshift, magnitude and luminosity distributions of our selected BOSS quasar and DLA samples.
%\yl{The $y$-axis labels are confusing in the right two panels.
%Why is green only in the left panel?} {\bf (Only QSOs have \textit{Magnitude} and \textit{Absolute magnitude}. DLAs are identified by absorption lines in QSO spectra.)}
%\yl{Thanks. I expected the y axis label be $dN/N\,dx$ with $x$ being the x
%variable of each panel.}
\label{fig:distribution}}
\end{figure*}

\section{Methodology}
\label{sec:method}

\subsection{Map construction and mask apodization}

To match the resolution of the CMB lensing convergence map,
we construct the quasar overdensity maps in the HEALPix format \citep{healpix} with $N_{\rm{side}} = 2048$, corresponding to 50,331,648 pixels for the whole sky:
\begin{equation}
    q_i = \frac{n_i-\bar{n}}{\bar{n}}
\end{equation}
where $i$ is the pixel number, $q_i$ is the quasar overdensity in the $i$-th pixel, and $n_i$ is the quasar number counts in each pixel with $\bar{n}$ being its mean value.
%{\bf should we show a figure here? Also, let's rephrase the size of each pixel here, and brief mention how 2048 related to the number of pixels.}

\begin{figure}[htbp]
    \centering
    \includegraphics[width=0.45\textwidth]{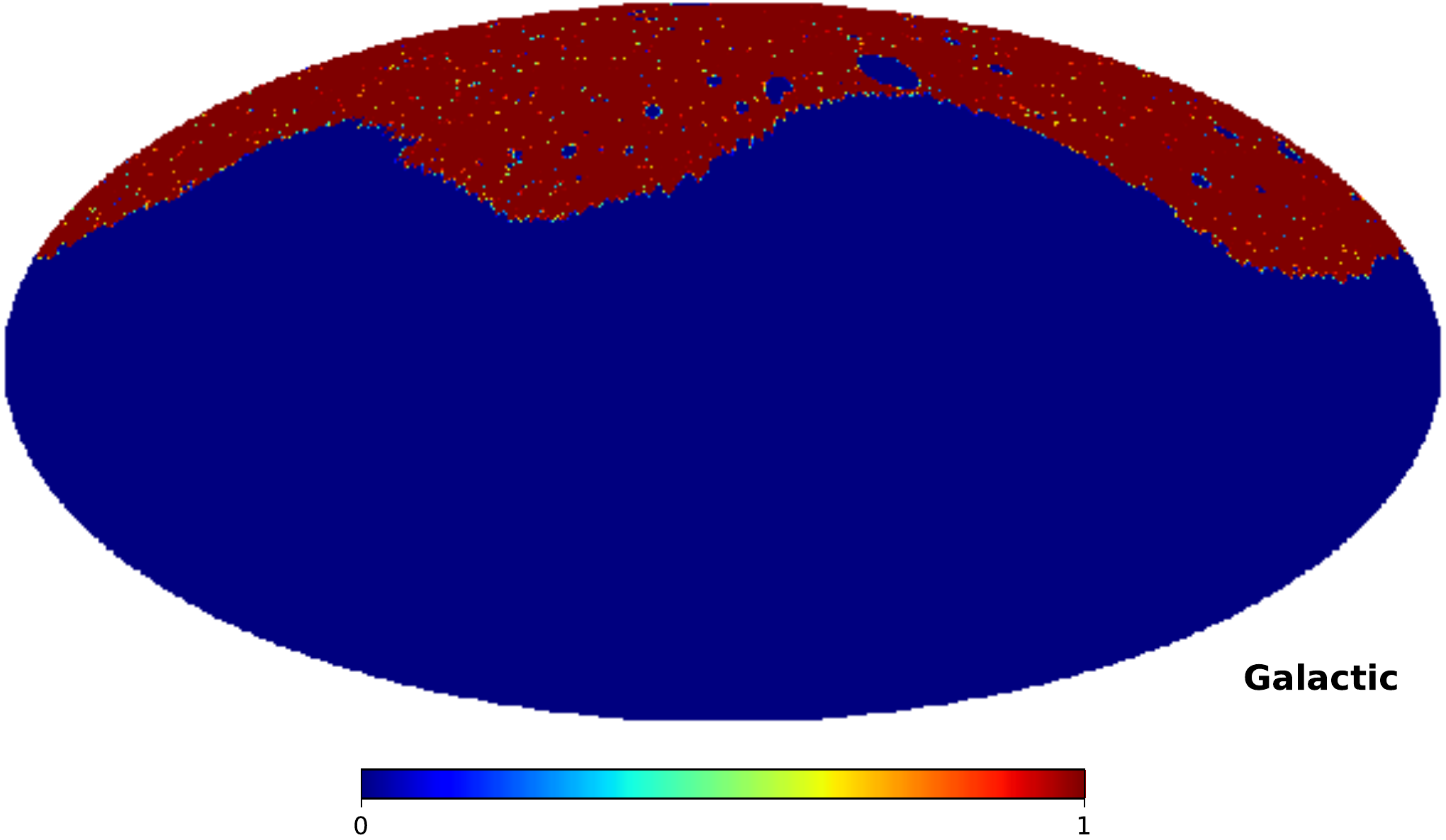}
    \caption{Apodized mask used to cross-correlated the quasar fields and CMB lensing convergence field. The mask has been smoothed by a Gaussian kernel of FWHM of 10 arcmin.}
    \label{fig:apodized_mask}
\end{figure}

Mask apodization is commonly used in the
cross-correlation calculation to avoid Gibbs-ringing induced by a sharp cut-off in a map during the harmonic transformation.
We give the brief procedures on mask construction and apodization
below, and visualize the apodized mask in Fig.~\ref{fig:apodized_mask}.
\begin{enumerate}[label=\roman*,align=CenterWithParen] \item We adopt
  the CMB lensing mask released by the Planck Collaboration as part of
  the \textit{Planck} 2018 CMB lensing products. As for the
    BOSS quasar masks, we lower the resolution of the quasar
    overdensity maps to $N_{\rm{side}}=32$ to identify the region
    covered by the SDSS survey. $N_{\rm{side}}=32$ corresponds to 12,288
    pixels or an angular resolution of 109.9 arcmin.
    This relatively low resolution is also chosen by \cite{Han} and
    \cite{Vielva2010}, to represent a continuous sky coverage allowed by
    the survey.
    We label the empty pixels 0 and the remaining 1 to construct the quasar mask. { A more accurate mask could be obtained from the random catalogs. We tried to use the CMASS random catalog provided by BOSS and get $\fsky^{\kappa q}$ = 0.184 (the fraction of the sky shared by the quasar overdensity map and the CMB lensing convergence map. See Eq.\ref{Cl_estimator} for more details), very close to the original $\fsky^{\kappa q}= 0.179$ by downgrading the quasar map, making negligible difference to the final results.}
  \item These masks are upgraded to $N_{\rm{side}}=2048$ again
%    \yl{unclear how. uniformly within each coarse pixels? In that
%    case I think we don't need to explain this step.} {\bf (Use the function \textit{healpy.pixelfunc.ud-grade(map-in,nside-out)}. I think explaining this step will make it easier to follow our work.)} \yl{I checked the docstring and the original healpix docs and couldn't find the algorithm there either. Let's keep it this way then. Thanks for explaining.}
before performing the cross-correlation, to match the resolution of quasar
    maps and the CMB lensing map.
    %{\bf one question: if we upgrade the masks to N=2048 from N=32,
    %then the mask will be different than that being directly made from
    %the data with $N_{\rm{side}}=2048$?} To avoid ringing issues in
    %spherical harmonic transform possibly induced by a sharp cut-off in
    %real-space,
  \item We apodize the masks by smoothing both the CMB mask and the
    quasar mask using a Gaussian kernel. To decide a proper kernel size, we generate 200 mocks based on a theoretical template, and apply masks smoothed by Gaussian kernels with FWHM of 10 arcmin, 30 arcmin and 1 degree, as well as a mask without smoothing. Then we calculate the mean squared error (MSE) of the residuals. A more detailed description for this procedure is given in Appendix \ref{apod}. We find that the mask smoothed by a Gaussian kernel with FWHM of 10
    arcmin produces the minimal residual and thus should be the optimal
    choice.
\end{enumerate}

\begin{table}[htbp]
    \centering
    \begin{tabular}{c|cccc}
    \hline
        FWHM & no smoothing & 10 arcmin & 30 arcmin & 1 deg \\
    \hline
        MSE/$\times 10^{-19}$ & 2.33 & 2.08 & 3.63 & 8.21 \\
    \hline
    \end{tabular}
    \caption{Comparison of different smoothing kernels, of width FWHM.
      MSE denotes the deviation of average binned spectra of the 200 Gaussian mocks from the theoretical template.
%   \yl{Mention again: apodization of masks, Gaussian kernel, smoothing
%      length. Unclear to me what is %``variance'' exactly (See comments
%      in text).}
    }
    \label{tab:kernel_size}
\end{table}

\subsection{Angular cross-power spectrum estimator}
\label{spectrum}

We compute the CMB lensing-quasar overdensity angular cross-power spectrum using a pseudo-$C_\ell$ estimator \citep{pseudo-estimator}:
\begin{equation}
  \hat{C}_{\ell}^{\kappa q}=\frac{1}{\fsky^{\kappa q}} \sum_{m=-\ell}^{\ell} \kappa_{\ell m}^{*} q_{\ell m}
  \label{Cl_estimator}
\end{equation}
where $\fsky^{\kappa q}$ is the fraction of the sky shared by the
quasar overdensity map and the CMB lensing convergence map. $\kappa_{\ell m}$ and $q_{\ell m}$ are the spherical harmonic transforms of the
CMB lensing convergence and the quasar overdensity maps,
respectively. { Note the 1/$\fsky^{\kappa q}$ estimator is known to be slightly biased \citep{Wandelt_2001}. However, while approximate, this should be good enough for our setup due to the noise level of the spectra.}
The transform can be computed readily by the anafast function within \texttt{healpy}\footnote{\url{https://github.com/healpy/healpy}}.
We bin both the BOSS QSO and QSO with DLAs cross-power spectra into 10 bands over the range of $100 < \ell < 1200$. 
{ We apply a low-$\ell$ scale cut with $\ell_\mathrm{min}=100$ because the cross correlation on $\ell < 100$ is deficient due to potential systematics shared
% to leave out the large scale regime where the
% Limber approximation used in Sec.~\ref{sub:theory} breaks down. 
%{such as the fluctuations on large scales induced by the incomplete sky coverage {\bf( Not a good guess: I think it is neglectable, because var induced by binary masks is $\sim 10^{-19}$ in Table 1)}}
%({\bf can we have some explanation of what ``some systematics" means?})
by the quasar overdensity map and the CMB lensing map. }
This deficit of power at $\ell<100$ %({\bf on the largest scale?})
is also found in \cite{Han,Anthony_2016,Ferraro_2016}, for which we have not found compelling explanations. { Some large scale systematics tracing the quasars may account for this \citep[e.g.][]{Geach_2019}.}%({\bf one systematics may be related to the incompleteness or selection function on this large scale.}).

Although \cite{Planck} has restricted the lensing auto-spectrum to the range of $8 \leq \ell \leq 400$ in the likelihood, %({\bf if the highest l is about 400 then why we can look up to l=1200?})
ensuring robustness of the reconstruction only over this range, we
perform a measurement on a higher multipole range with
$\ell_{\rm{max}}=1200$ \footnote{We perform the same analysis described in this paper for different $\ell_{\rm max}$ and number of bins. For BOSS QSOs, the spectrum within 10 bins with $\ell_{\rm max}=1200$ yields the minimum $\chi^2$. Due to the low SNR of our DLA signal, we can only give a rough estimate for the upper limit of DLA halo mass and thus we simply choose the same $\ell$ range and number of bins for QSOs with DLAs.}, expecting both noise and systematics to be
subdominant over this range as discussed in \cite{G_2016}.

\subsection{Cross-power spectrum covariance matrix}\label{cov}
\label{error}

There are two common ways to estimate the statistical errors on the
cross-power spectrum.
One is through analytical calculation assuming both fields are Gaussian
\citep{errorbar,Hivon2002,EG2004}:

\begin{equation}
  \frac{1}{\sigma_i^{2}(A)} =
  \sum_{l_{\min }(A)<l<l_{\max }(A)}
  \frac{\fsky^{\kappa q}(2 l+1)}{\left(C_{l}^{\kappa q}\right)^{2}+C_{l}^{\kappa \kappa} C_{l}^{q q}}
  \label{eq:sigma}
\end{equation}
%
%\yl{This equation looks strange.
%  I would write something like
%  \begin{equation}
%    \sigma^{2}_i =
%    \frac{\sum_{\ell_i \leq \ell < %\ell_{i+1}}
%      (2\ell+1) \bigl[ \left(C_\ell^{\kappa %q}\right)^2
%        + C_\ell^{\kappa \kappa} C_\ell^{q %q} \bigr]}
%    {\Bigl[ \sum_{\ell_i \leq \ell < %\ell_{i+1}}
%      \fsky^{\kappa q} (2\ell+1) \Bigr]^2},
%    \label{eq:sigma}
%  \end{equation}
%  \yl{Have you compared this equation with the literature?}{\bf (The 2 equations seem different?)}
  where $\sigma_i^2$ is the error of the cross power in the $i$-th bin. $C_\ell^{\kappa \kappa}$ and $C_\ell^{qq}$ are the expected CMB lensing and quasar auto-power spectra, including both signal and noise.
The estimators of the auto-power spectra are
\begin{align}
  \hat{C}_{\ell}^{\kappa \kappa} &= \frac{1}{\fsky^{\kappa}(2 \ell+1)}
    \sum_{m=-\ell}^{\ell}\left|\kappa_{\ell m}\right|^{2}, \\
  \hat{C}_{\ell}^{q q} &= \frac{1}{\fsky^{q}(2 \ell+1)}
    \sum_{m=-\ell}^{l}\left|q_{\ell m}\right|^{2}.
\end{align}

Alternatively one can get the error as the diagonal of
the covariance matrix of $\hat{C}^{\kappa q}_\ell$.
To estimate the latter, we use 300 CMB lensing simulations
%\yl{300 mock lensing maps?}
%{\bf (It is called CMB lensing simulations in previous works, such as
%\citet{Han})}
and follow the steps described in \S~\ref{spectrum} to calculate their
cross-power spectra with our BOSS quasar sample and selected quasars
with DLAs.
The 300 simulations of CMB lensing convergence field we adopt are part
of the \textit{Planck} 2018 release, based on MV estimate from
temperature and polarization of all 300 \textit{Planck} Full Focal Plane
(FFP10) simulations \citep{Planck} \footnote{The Full Focal Plane
simulations are used to generate multiple mission realizations, and
these maps incorporate the dominant instrumental, scanning, and data
analysis effects \citep{FFP10}.}.

The covariance matrix element between the $i$-th and the $j$-th bin,
$\mathcal{C}_{ij}$,  can be estimated by
\begin{equation}
  \mathcal{C}_{ij} = \frac{1}{N-1} \sum_{n=1}^{N}
  (C_{i,n}-\bar{C_i})(C_{j,n}-\bar{C_j})
\end{equation}
where $N=300$ is the total number of CMB lensing simulations, and $C_{i,n}$ denotes the $i$-th bin of the cross-power spectrum between the $n$-th mock CMB lensing simulations and our quasar sample. 
The error of each bin can be obtained from the square root of the
variance, i.e.\ the diagonal of the covariance matrix:
\begin{equation}\label{eq:err}
    \sigma_i = \sqrt{\mathcal{C}_{ii}}.
\end{equation}

\begin{figure}[htbp]
    \centering
\includegraphics[width=0.6\textwidth]{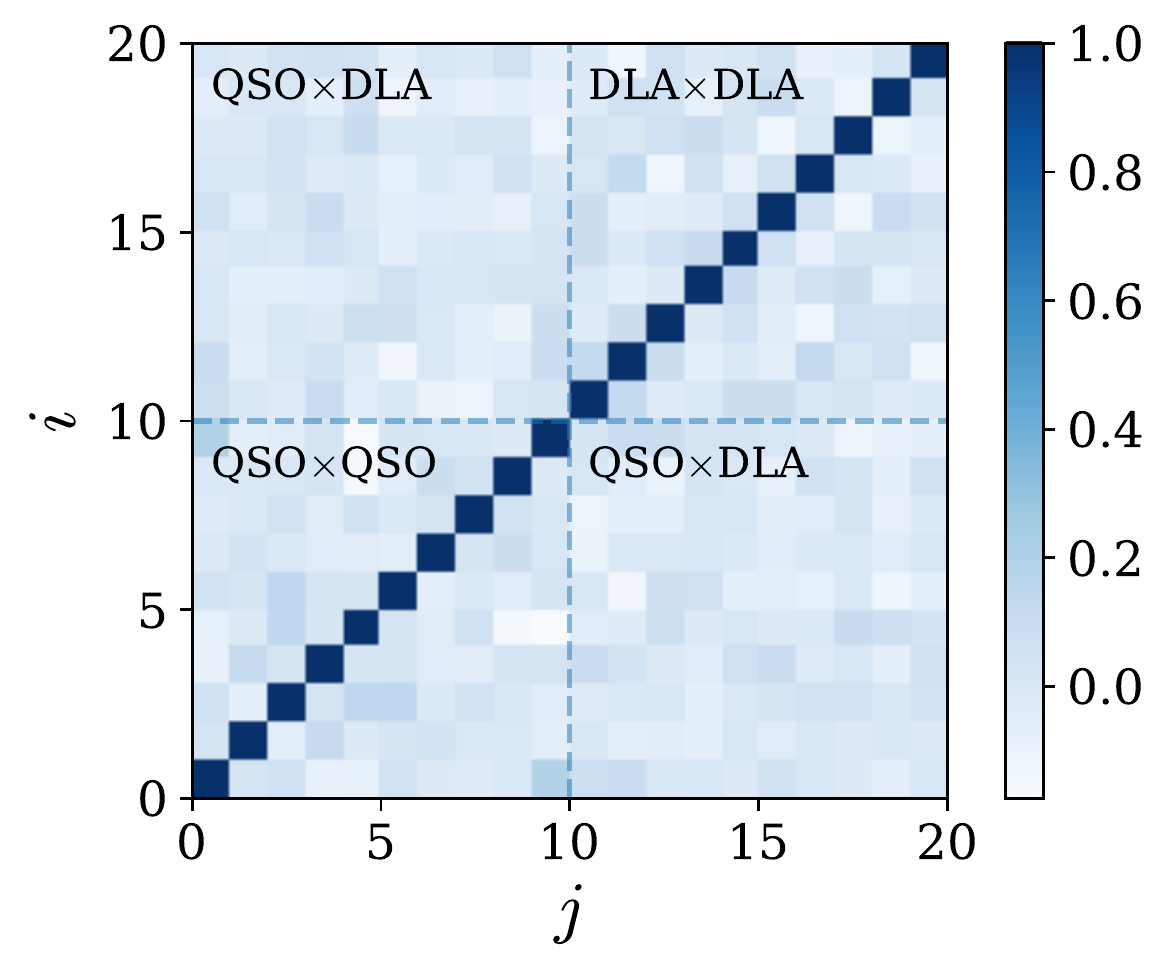}
    \caption{The correlation matrix for the BOSS QSOs and QSOs with DLAs. Here $i=1-10$ denote bins in the spectrum of the BOSS QSOs and $i=11-20$ denote those of QSOs with DLAs.}
    \label{fig:cov}
\end{figure}

With the covariance matrix, it is convenient to compute the
correlation matrix, which describes the linear correlation between pairs
of bins:
\begin{equation}
  r_{ij} = \frac{\mathcal{C}_{ij}}{\sqrt{\mathcal{C}_{ii}\mathcal{C}_{jj}}},
\end{equation}
with $-1 \leq r_{ij} \leq 1$.
Two bins are tightly positively correlated when $r$ is close to 1,
negatively correlated when $r$ is close to $-1$, and linearly
uncorrelated when $r = 0$.
We show the correlation matrices of lensing$\times$quasar and
lensing$\times$DLA in Fig.\ref{fig:cov}. { As one can see, the off-diagonal blocks can be negligible in this case. First of all, the bins are fairly wide, and so the bin-bin covariance should be pretty small. Then, due to the big difference in the number count of the two catalogs,  the overlapping part of the two catalogs is so small that their covariance can be ignored.}

%({\bf also, we need to explicitly give the error expression using the covariance matrix method, say $\sigma=1/\sqrt{r_{\rm{ij}}}$ just same as Eq.(2)}).

We adopt Eq.\ref{eq:err} in the bias measurement, and Eq.\ref{eq:sigma} in the null test (\S~\ref{sec:null}).
%\yl{Should there be some explanation? Have you compared the two?} {\bf (just for convenience... A null test using 300 CMB lensing simulations is a little expensive. I think a null test with the analytical Gaussian estimator is enough.)}

%% \begin{figure}[htbp]
%%    \centering
%%    \gridline{
%%        \fig{cov_q_10arcmin.pdf}{0.35\textwidth}{the BOSS subset}
%%        \fig{cov_dla_10arcmin.pdf}{0.35\textwidth}{QSOs with DLAs}
%%          }
%%     
%%          \caption{The correlation matrices for the BOSS QSOs $\times$ %% CMB lensing (left) and selected quasars with DLAs $\times$ CMB lensing %% (right). %\yl{These seem to be auto-correlation. What about cross with %% lensing?}{\bf (These are auto-correlations of lensing signal around the %% 2 QSO catalogue. We think the BOSS QSO $\times$ DLA QSO correlation  matrix is not necessary because it won't have a big impact on our parameter constraints.)}
         % \yl{Sure. But what about cross \textbf{with lensing},
         %   $C^{\kappa q}$? Maybe I am confused.
        % What is exactly shown here? Sound like $r^{qq}$ by the caption title.}
%%            }
    
%% \end{figure}

\subsection{Analytical Model}
\label{sub:theory}

\subsubsection{Measurement of quasar bias and DLA bias} \label{measure b}

%\yl{why subsubsection and why only one?}
%{\bf (Firstly describe the basic theoretical correlation curve; then
%introduce the fiducial bias model and the scaling parameter; and derive
%halo mass from bias.)}
%\yl{So 3.5 should be 3.4.2 and 3.6 likewise?}

%\yl{Given that we cite Loverde, we should probably replace $\ell$ with
%$\ell+1/2$, at least in the equations.}

%\yl{It'd be better to introduce equation and relevant physics in logical
%order: $\kappa$, $W$; $q$, $f\&g$; and finally $C$.}
The CMB lensing convergence is defined as a weighted projection of the matter overdensity in the direction along the line of sight $\hat{n}$ \citep{Lewis}:
\begin{equation}
\kappa(\hat{n})=\int^{z_{\rm{CMB}}}_{0}dzW(z)\delta_m \left(\chi(z)\hat{n},z\right)
\end{equation}
$\delta_m(\chi(z)\hat{n},z)$ is the matter overdensity at redshift $z$ in the direction $\hat{n}$, $W(z)$ is the CMB lensing kernel:
\begin{equation}
 W(z)=\frac{3 H_{0}^{2} \Omega_{m, 0}}{2 c H(z)}(1+z) \chi(z)\left(1-\frac{\chi(z)}{\chi_{\mathrm{CMB}}}\right)
 \end{equation}
where $H_0$ is the current Hubble parameter, $H(z)$ is the Hubble parameter at redshift $z$, $\Omega_{m,0}$ is the current matter density, $c$ is the speed
of light, $\chi(z)$ is the comoving distance to redshift $z$ and $\chi_{\rm{CMB}}$ is the comoving distance to the last scattering surface at redshift $z_{\rm{CMB}}\approx1100$.

To relate the quasar overdensity field to the matter overdensity field, the projected surface density of quasars is given by \citep{Peiris_2000}:
\begin{equation}
 q(\hat{n})=\int_{0}^{z_{\mathrm{CMB}}} d z f(z) \delta_{m}(\chi(z) \hat{n}, z)
\end{equation}
where $f(z)$ is its window function:
\begin{subequations}\label{eq:fz}
\renewcommand{\theequation}{\theparentequation.\arabic{equation}}
\begin{equation}
f(z)=\frac{b(z) d N / d z}{\left(\int d z^{\prime} \frac{d N}{d z^{\prime}}\right)}+\frac{3}{2 H(z)} \Omega_{0} H_{0}^{2}(1+z) g(z)(5 s-2)
\end{equation}
\begin{equation}
g(z)=\frac{\chi(z)}{c}\int_{z}^{z_{CMB}} d z^{\prime} \left(1-\frac{\chi(z)}{\chi(z^\prime)}\right)\frac{d N}{d z^{\prime}}
\end{equation}
\end{subequations}
The first term is the normalized, bias-weighted redshift distribution of the quasars; the second term is the magnification bias induced by the change in the density of sources due to lensing magnification,
where $s \equiv d \log _{10} N / d m$ is the  response of the number density to magnification bias. Due to the multiple selection criteria, we cannot simply histogram the sample to determine $s$. Instead, following \cite{Simone_mag_bias}, % we make all the quasars in the BOSS subset fainter by 0.1 magnitudes and those in the DLA catalog fainter by 0.01 {\bf why choose 0.01 for QSO with DLA but 0.1 for QSO?} magnitudes, re-apply the same selection criteria and measure the changes in their number counts to determine $s$. {Answer:}The shifting in magnitudes is different due to the different sizes of the two samples. As a result,
we get $s \approx 0.078$ for the BOSS subset and $s\approx 0.26$ for the quasars with DLAs, %both making the magnification bias significantly small compared to the intrinsic clustering of quasars, { and thus we just neglect the uncertainty induced by the magnification bias.}{\bf then what? negliect the second term? we need to explicity say the consequence here. } ({\bf also, should we only compare $s$ to $b$, or should compare the whole first term to second term to neglect the second term? }{ Answer:the 2nd term v.s. the 1st term}).
and just neglect the uncertainty induced by the magnification bias.

For quasars with DLAs, DLAs contribute to the cross-correlation between CMB lensing map and the quasar overdensity. Simply assuming the weak lensing signal of CMB photons is the summation of the quasar and its foreground DLA contribution, the window function in Eq.\ref{eq:fz} should be written as :
\begin{equation}\label{fz_dla}
f(z)=\frac{\bar{n}_{\rm DLA}\bdla\cdot dN_{\rm{DLA}}/dz}{\left(\int dz^{\prime} \frac{dN_{\rm{DLA}}}{dz^{\prime}}\right)} + \frac{\bqso\cdot dN_{\rm{QSO}}/dz}{\left(\int dz^{\prime} \frac{dN_{\rm{QSO}}}{dz^{\prime}}\right)}
+\frac{3}{2 H(z)} \Omega_{0} H_{0}^{2}(1+z) g(z)(5 s-2)
\end{equation}
$\bar{n}_{\rm DLA}$ is the effective number count of DLAs for a single sightline. In this case it should be $N_{\rm DLA}/N_{\rm QSO}\approx1.17$. { Note that here $N_{QSO}$ denotes the number of background quasars which have DLAs found in their spectra, different from $N$ in Eq.\ref{eq:fz} referring to the number of the full quasar sample.}

In a flat universe and under the Limber approximation
\citep{Limber1953}, the quasar-CMB lensing convergence angular
cross-power spectrum is given by \citep{Lewis,Sherwin_2012}:
\begin{equation}\label{Cl_theory}
C_{\ell}^{\kappa q}=\int \frac{d z}{c} \frac{H(z)}{\chi^{2}(z)} W(z) f(z) P_{m m}\left(k=\frac{\ell}{\chi(z)}, z\right)
\end{equation}
where  and
$P_{mm}(k, z)$ is the 3D matter power spectrum.

\begin{table}[h]
    \centering
\begin{tabular}{cccccccc}
\hline$\Delta z$ & $\Delta M_{i}$ & $\bar{z}$ & No. of quasars & $b_{Q}$ & $\chi_{\text {red }}^{2}$ \\%& $r_{\text {o}}\left(h^{-1} \mathrm{Mpc}\right)$ & $\chi_{\text {red }}^{2}$ \\
\hline $2.20 \leq z \leq 2.80$ & $-28.74 \leq M_{i} \leq-23.78$ & 2.434 & 55826 & $3.54 \pm 0.10$ & 1.06 \\ %& $8.12 \pm 0.22$ & 0.25 \\
\hline $2.20 \leq z<2.384$ & $-28.70 \leq M_{i} \leq-23.95$ & 2.297 & 24667 & $3.69 \pm 0.11$ & 1.55 \\%& $8.68 \pm 0.35$ & 0.45 \\
$2.384 \leq z<2.643$ & $-28.74 \leq M_{i} \leq-24.11$ & 2.497 & 24493 & $3.55 \pm 0.15$ & 0.61 \\%& $8.42 \pm 0.54$ & 0.26 \\
\hline $2.643 \leq z \leq 3.40$ & $-29.31 \leq M_{i} \leq-24.40$ & 2.971 & 24724 & $3.57 \pm 0.09$ & 0.66 \\%& $7.59 \pm 0.66$ & 0.33 \\
\hline $2.20 \leq z \leq 2.80$ & $-28.74 \leq M_{i}<-26.19$ & 2.456 & 18477 & $3.69 \pm 0.10$ & 2.19 \\%& $8.62 \pm 0.27$ & 0.94 \\
$2.20 \leq z \leq 2.80$ & $-26.19 \leq M_{i}<-25.36$ & 2.436 & 18790 & $3.56 \pm 0.13$ & 1.71 \\%& $7.94 \pm 0.41$ & 1.63 \\
$2.20 \leq z \leq 2.80$ & $-25.36 \leq M_{i} \leq-23.78$ & 2.411 & 18559 & $3.81 \pm 0.19$ & 0.4 \\%& $8.29 \pm 0.36$ & 3.29 \\
\hline
\end{tabular}
    \caption{Clustering results for NGC-CORE sample and subsamples in \cite{BOSS_auto} (Table 5 in their paper). The first four columns are redshift and absolute magnitude range, the average redshift and the total number of NGC quasars. Columns 5, 6 are the best-fitting bias values and the reduced $\chi^2$ (over 7-DoF/9-DoF for the main sample/subsamples). }%The last two columns are the correlation lengths and their reduced $\chi^2$ (over 9-DoF/8-DoF for the main sample/subsamples).}
    \label{tab:auto-corr}
\end{table}

\subsubsection{The fiducial bias model}

We adopt the quasar bias measured in \cite{BOSS_auto} as our fiducial bias model. They report one of the most precise measurements of quasar bias for the BOSS sample using the auto-correlation approach. The quasar bias (see Table \ref{tab:auto-corr}, repeating Table 5 in \citet{BOSS_auto}) has no significant upward trend as $z$ increases. We use $\bar{b}=3.60$, the average of their $\bqso$, as the fiducial value, and assume a scaling parameter $a$ which depicts the level of deviation from the fiducial $b_{\rm{fid}}$:
\begin{equation}\label{scaling_a}
    b = a\cdot b_{\rm{fid}}
\end{equation}

As a test, we also adopt the bias-redshift model derived in \cite{Laurent2017} to guarantee the robustness of our analysis. More details are shown in Appendix \ref{Laurent17}.

\subsubsection{From bias to halo mass}
\label{sub:b2M}

We use the fitting formula in \cite{Tinker10} to relate the bias to the
peak height of the linear density field $\nu$
\citep{Press1974,Sheth2001}:
\begin{equation}
\begin{aligned}
\nu&=\frac{\delta_{c}}{\sigma(M_{\Delta})}\\
b(\nu)&=1-A \frac{\nu^{\alpha}}{\nu^{\alpha}+\delta_{c}^{\alpha}}+B \nu^{\beta}+C \nu^{\gamma}
\end{aligned}
\end{equation}
where $A=1.0+0.24 y \exp \left[-(4 / y)^{4}\right]$, $\alpha=0.44y-0.88$, $B=0.183$, $\beta=1.5$,
$C = 0.019+0.107 y+0.19 \exp \left[-(4 / y)^{4}\right]$ and $\gamma=2.4$ if we assume $y=log_{10}\Delta$. $\delta_c=1.686$ is the critical overdensity for collapse, and $\sigma(M_\Delta)$ is the linear matter variance on the Lagrangian scale of the halo.

In this work we assume $\Delta=200$ and thus $M$ refers to the total mass within the radius $r_{200}$ at which the enclosed mass density is 200 times the average matter density of the Universe:
\begin{equation}M_{200}=\frac{800 \pi}{3} \rho_{\text{m}}(z) r_{200}^{3}.\end{equation}
%
%%\texttt{massFromPeakHeight} in 
\texttt{COLOSSUS}\footnote{\url{https://bdiemer.bitbucket.io/colossus/}} \citep{colossus} is a convenient tool to derive $M$ from the peak height $\nu$. Since the relation between bias and $\nu$ is slightly complicated, we invert the mass-bias relation and find the roots using \texttt{scipy}.
%by \texttt{fsolve} in \texttt{scipy.optimize}.
%{\bf what is the relation between $\nu$ and $b$?}

\section{Results}\label{sec:results}

In this section we present our measurements of the quasar and DLA
biases.
We use the quasar bias measured from the BOSS sample as a prior to help
us constrain the DLA bias.

\subsection{Quasar bias}\label{sec:b_qso}

First, we cross-correlate the BOSS quasar sample with the CMB lensing
map to { get the
maximum-likelihood estimate of $\bqso$} using Eq.\ref{Cl_estimator} and
Eq.\ref{Cl_theory}.
{ We reweight the redshift distribution of quasars in the BOSS subset following \cite{JCAP_DLA}, so that it matches that of the QSOs with DLAs. The reweighting is to ensure the assumption that quasars in the two catalogs have similar bias. In fact, in this case reweighting the sample makes negligible difference to the final results.}
We run \texttt{CAMB} \citep{Lewis_2000} to calculate the nonlinear
matter power spectrum using HALOFIT \citep{Smith_2003, Takahashi_2012} .
%% We apply \texttt{curve\_fit} in
%% \texttt{scipy.optimize}\footnote{\url{https://docs.scipy.org/doc/scipy}}
%% to get the best fitting value.
In Fig.\ \ref{fig:spectrum}, we present the cross-correlation (black
point with error bar).
We use the standard deviation of the fitting result as a proxy of its
uncertainty, which can be derived from the resulted covariance from the
curve fitting.

With the parametrization in Eq.\ref{scaling_a}, we estimate the
parameter $a = \BOSSaqso$ using Eq.\ref{eq:fz}.
%\yl{Only Eq.\ref{eq:fz} right? Eq.\ref{fz_dla} is for the next subsection?}
This corresponds to $\bqso= a\cdot b_{\rm{fid}}=\BOSSbqso$.
Following \S~\ref{sub:b2M}, the bias value corresponds to a
characteristic halo mass of $\log(M/M_\odot
h^{-1})=\BOSSmqso$ at a median redshift of $z_{\rm
m}\approx2.51$.

\subsection{DLA bias}

We cross-correlate the \textit{Planck} 2018 CMB lensing map with the
17,774 quasars which have DLAs in their sightlines.
If we ignore the impact of DLAs and follow the steps introduced in
\S~\ref{measure b}, we obtain $a=1.00\pm0.55$, deviating from the value
of $\bqso$ estimated in \S~\ref{sec:b_qso}.
Therefore, the contribution of DLAs in sightlines of quasars is not negligible.

We measure $\bqso$ and $\bdla$ with the Markov Chain Monte Carlo (MCMC)
using \texttt{emcee\footnote{\url{https://github.com/dfm/emcee}}}
\citep{emcee}.
Since the luminosity distribution of the BOSS subset and the selected
quasars with DLAs are not significantly different (see
Fig.\ref{fig:distribution}
%\yl{wrong link?}
), we assume that $\bqso$ of quasars
with DLAs approximates that of the BOSS QSO subset.
Then, we assume the DLA bias $\bdla$ is a constant over the selected redshift range at $z=2.2- 3.4$,
with a prior uniform distribution over $0<\bdla<15$ %({\bf what is the prior value of $\bdla$?}),
and further assume a Gaussian distribution of $a$ as derived in \S~\ref{sec:b_qso}, with { the central value $\mu=0.71$ and the standard deviation $\sigma=0.19$ as its prior distribution}:
\begin{equation}
    p(a)\sim N(\mu,\sigma^2)
\end{equation}
The likelihood function is:
{
\begin{equation}
    p(\textbf{C}|a,\bdla)=\frac{1}{|\Sigma|^{\frac{1}{2}} (2 \pi)^{\frac{n}{2}}} e^{-\frac{1}{2}\left({\bf {\hat C}}-{\bf C}\right)^T{\bf \Sigma}^{-1}\left({\bf{\hat{C}}}-{\bf C}\right)}
\end{equation}
}
where $\hat{\bm C}$ is the data vector of the binned data points,  $\bm C$ is the data vector of calculated theoretical model for the bin, and
$\bm \Sigma$ is covariance matrix as shown in Fig.\ref{fig:cov}.
%\yl{I would say $\hat{\bm{C}}$ is the data vector of $\hat C_i$}.

\begin{figure}
\gridline{
        \fig{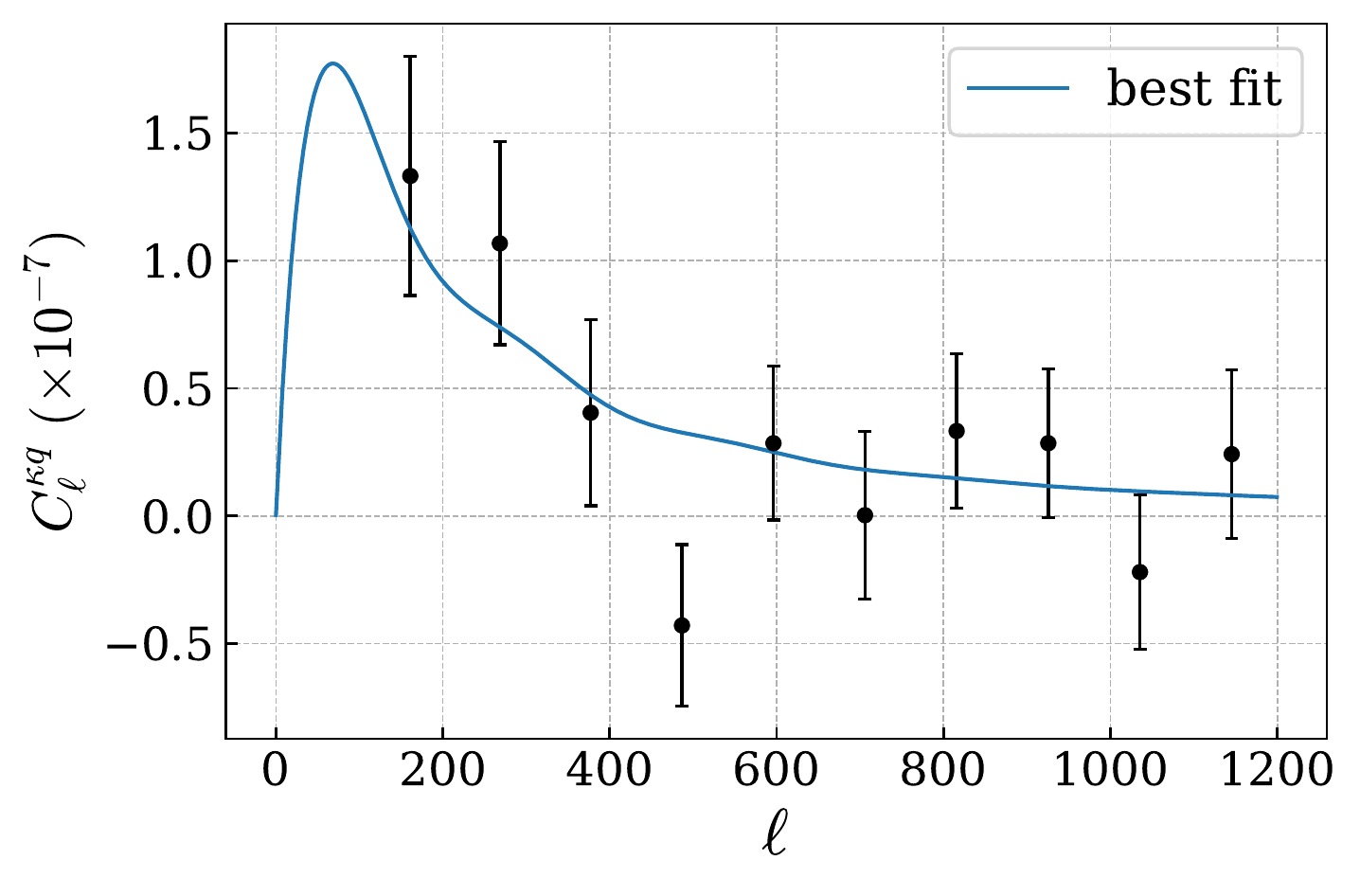}{0.45\textwidth}{the BOSS subset}
        \fig{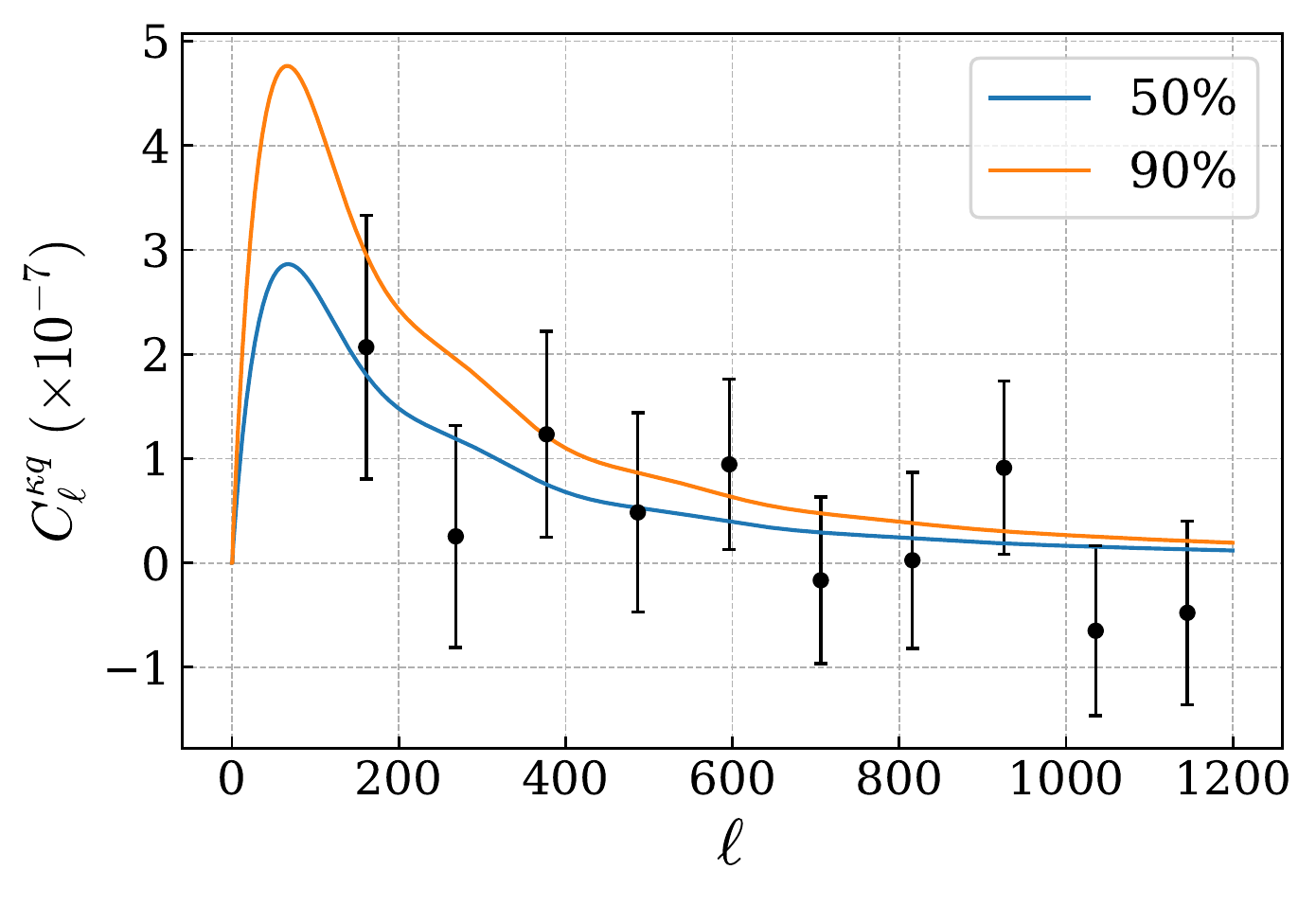}{0.45\textwidth}{QSOs with DLAs}
          }
\caption{The CMB lensing-quasar overdensity angular cross-power spectrum for the BOSS subset (the left panel) and the QSOs with DLAs (the right panel) over the redshift range of $2.2<z<3.4$. In the left panel, the solid line is the best-fit theoretical curve. In the right panel, the two solid lines denote the theoretical cross-power spectra at 50\% and 90\% upper limits of $\bdla$.
}\label{fig:spectrum}
\end{figure}

\begin{figure}
\gridline{
        \fig{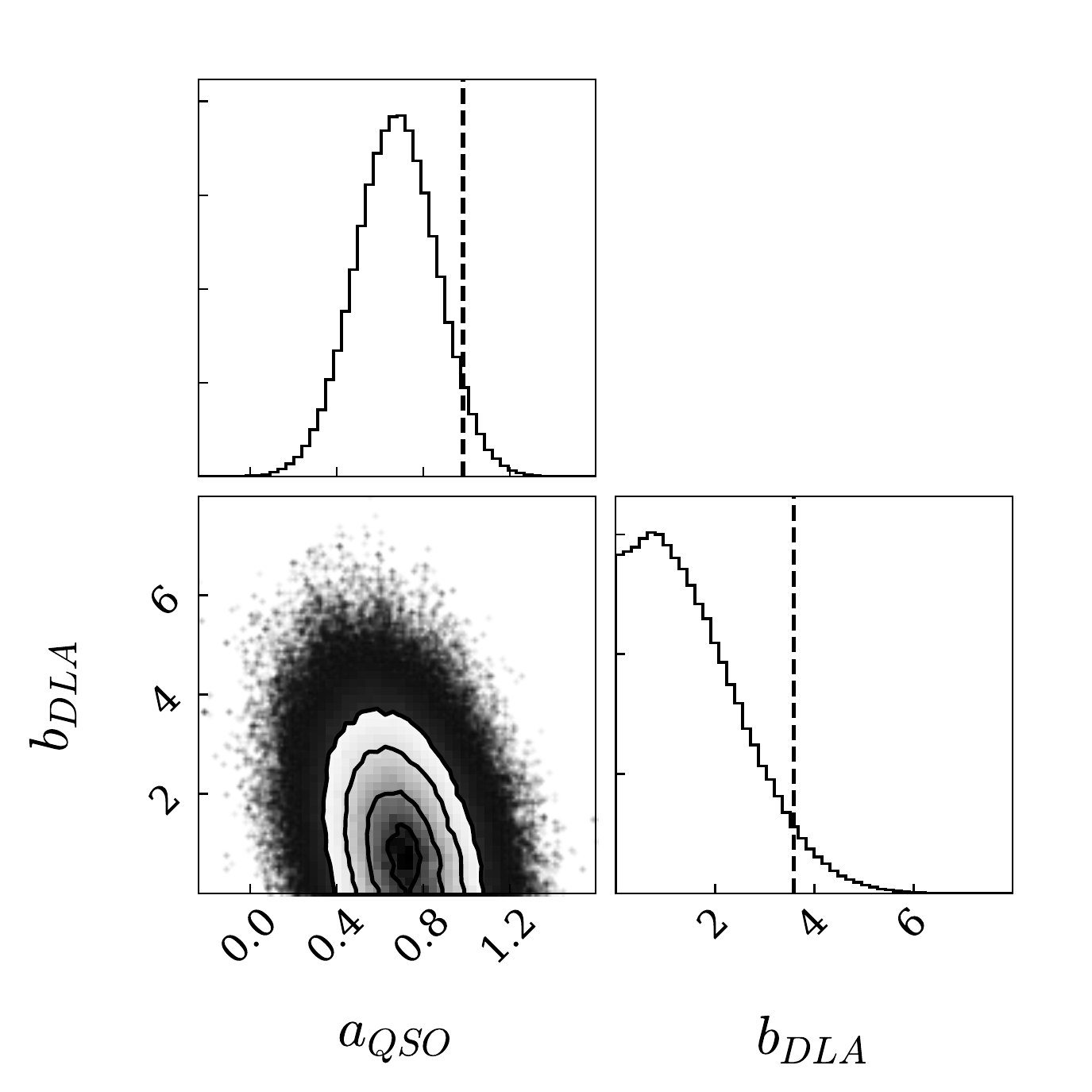}{0.45\textwidth}{}
          }

\caption{ One and two dimensional projections of the posterior probability distributions of $a$ and $\bdla$. The dashed lines denote the 90\% upper limit, i.e. 90\% of the MCMC samples in the marginalized distributions fall into its left region.
}
\label{fig:mcmc_plot}
\end{figure}

We have $a=\DLAaqso$, and $\bdla=\DLAbdla$ corresponding to a characteristic DLA halo mass of $\log(M/M_\odot h^{-1})=\DLAmdla$ at a median redshift of $z_{\rm m}\approx2.30$. Despite the large uncertainty, the result provides a good constraint on the upper limit of DLA halo mass, yielding $\bdla\leq\upbdla$ and log$(M/M_\odot h^{-1})\leq \upmdla$ at a confidence level of $90\%$. The results are shown in Fig.\ref{fig:spectrum} and Fig.\ref{fig:mcmc_plot}.

{ We have also tried to simultaneously modelling on the two catalogs, assuming that their $\bqso$ are exactly the same. The results shows that $a_{\rm QSO}=0.67\pm0.19$, $\bdla=1.57^{+1.33}_{-1.02}$ and the 90\% upper limit of $\bdla$ is 3.30, corresponding to a halo mass of log$(M/M_\odot h^{-1})\leq12.38$. This is consistent with our analysis.}

%The blue solid lines denote that 50\% of the MCMC samples in the marginalized distributions fall into its left region, and the dashed lines denote 16\% and 84\%.

\section{Null test}\label{sec:null}
\begin{figure}[htbp]
    \centering
\gridline{
\fig{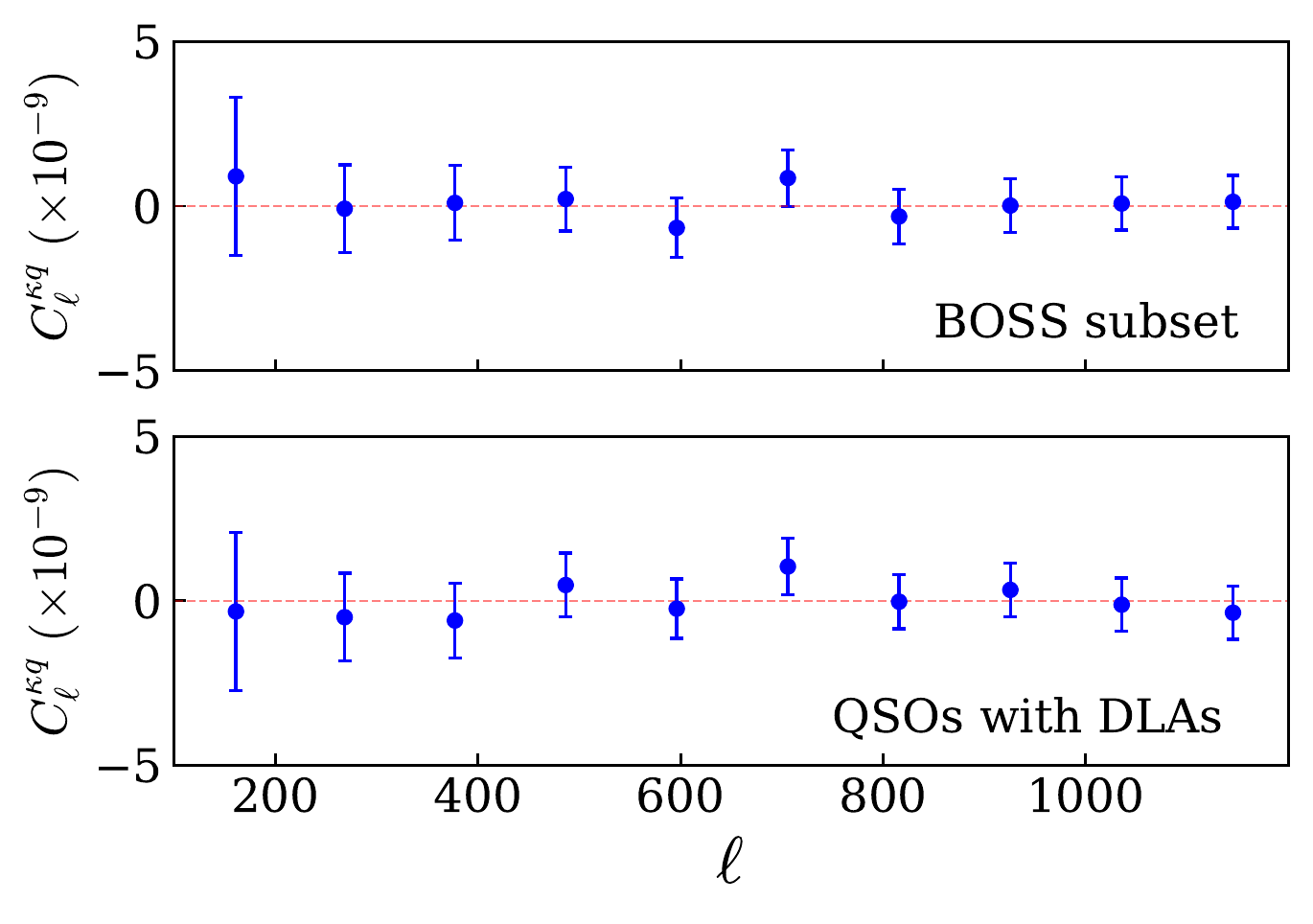}{0.45\textwidth}{(a)}
\fig{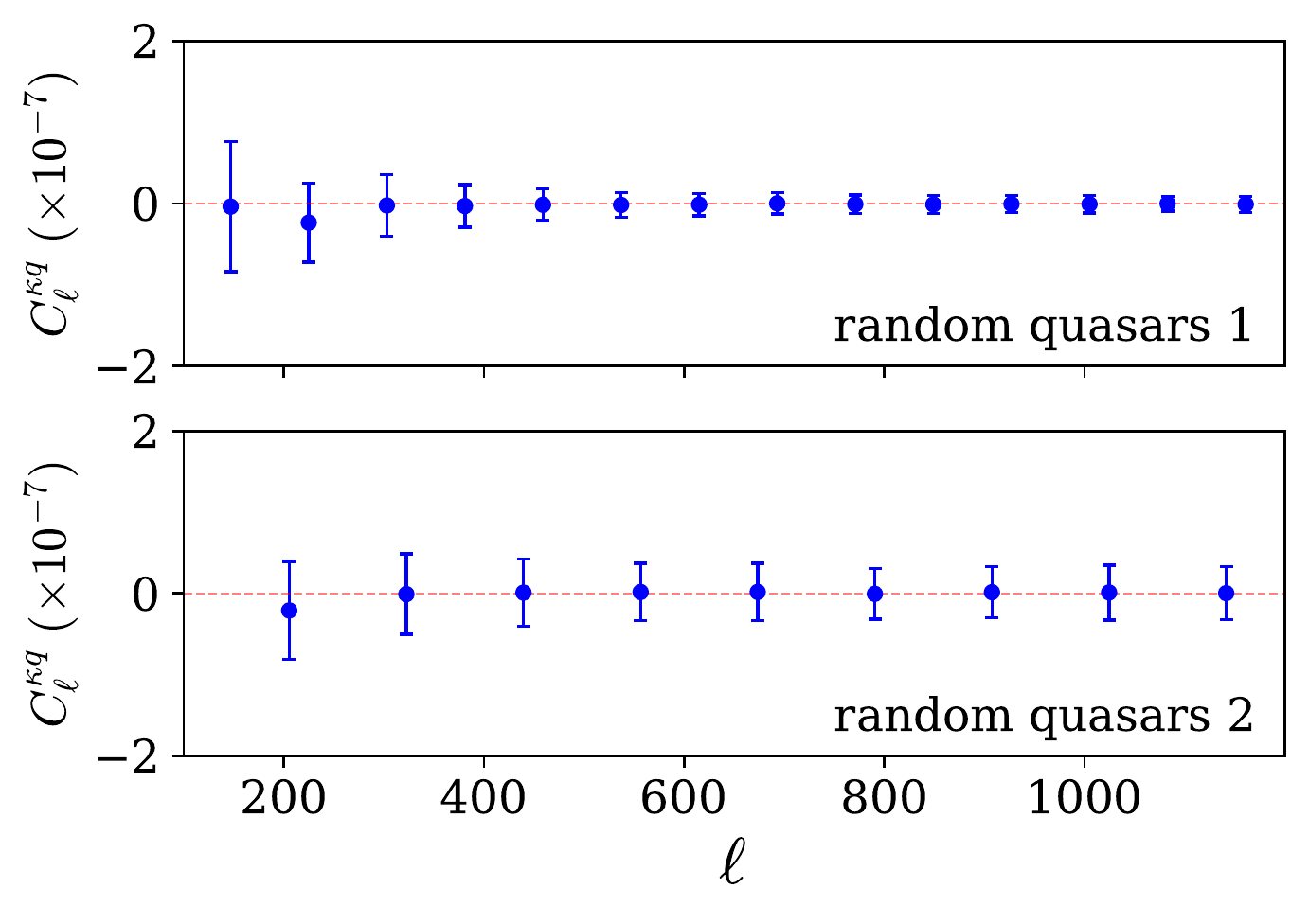}{0.45\textwidth}{(b)}
}
    \caption{\textit{Left:} Results of null tests by cross-correlating $0^\circ\sim100^\circ$ of CMB lensing map and $120^\circ\sim220^\circ$ of the 2 quasar maps. The error bars are obtained by Eq.\ref{eq:sigma}. \textit{Right:} Results of null tests by cross-correlating the \textit{Planck} FFP10 simulations and the 2 randomly-positioned quasar maps. Here \textit{random quasars 1} refers to a catalog of 300,000 quasars with random positions and \textit{random quasars 2} refers to a catalog of 100,000 quasars with random positions. The error bars are obtained by Eq.\ref{eq:err}.}
    \label{fig:null_test}
\end{figure}

%\yl{Can this be part of previous section?}
We check our result by a simple null test \citep{Sherwin_2012}, calculating the cross-power spectrum between the CMB lensing convergence map on one part of the sky and the quasar map on another part of the sky. The galactic longitude of the selected part of CMB lensing map ranges from $0^\circ$ to $100^\circ$, and that of the 2 quasar maps ranges from $120^\circ$ to $220^\circ$. The error bar of the spectra is calculated by the analytical Gaussian error estimator described in \S~\ref{error}. Almost all the bins in their cross-power spectra fall within 1$\sigma$ of null, as shown in Fig.\ref{fig:null_test}, yielding $a=0.004\pm0.007$ for the BOSS subset and $a=0.001\pm0.007$ for the selected quasars with DLAs.

{ We also correlate the 300 \textit{Planck} FFP10 simulations with randomly-positioned quasars, and get the average of the 300 spectra as an additional null-test, yielding $a=-0.046\pm 0.131$ for 300,000 randomly-positioned quasars within 15 bins and $a=-0.046\pm0.296$ for the other 100,000 quasars within 10 bins. }

\section{Discussion}\label{sec:discuss}

Our measurement of bias for the BOSS subset ($a=\BOSSaqso$,
$\bqso=\BOSSbqso$), albeit with large uncertainty, is lower than the
result in \cite{BOSS_auto} measured by two-point auto-correlation
function. % A lower quasar bias is also reported in \cite{JCAP_DLA}.
For the DLA-CMB lensing cross-correlation, \citet{JCAP_DLA} pioneered using the CMB lensing convergence map to constrain the DLA bias. They used the \textit{Planck} 2015 CMB lensing map and 2 DLA catalogs labelled as N12 \citep{N12} and G16 \citep{G16}, both created from SDSS-\uppercase\expandafter{\romannumeral3} DR12. They yield constraints on the QSO bias of $2.90\pm0.69$ ($2.57\pm0.52$) and DLA bias of $2.66\pm0.93$ ($1.92\pm0.69$) for their N12 (G16) sample by simultaneously modelling the DLA+QSO and QSO-only measurement.
Despite large uncertainties, our DLA bias has a 1.5-$\sigma$ lower value than that of \citet{JCAP_DLA}.
Comparing with \citet{JCAP_DLA}, our study has several updates:
(1) we adopt \textit{Planck} 2018 CMB lensing convergence map and an
updated DLA catalogue produced by a CNN model, currently the
state-of-the-art finding algorithm for DLAs;
(2) we apply similar sample selection criteria as described in
\citet{BOSS_auto} so that the QSO sample we used for measuring $\bqso$
is more comparable to that of the auto-correlation results;
(3) We apply the same sample selection criteria to both the BOSS quasars
and the quasars with DLAs, including removing radio-loud sample, so that
the measured $\bqso$ of the BOSS subset is comparable to $\bqso$ of the
selected quasars with DLAs;
(4) We have an updated noise calculations, quantitatively considering
the different apodization size and the tSZ effects (see Appendix
\ref{non-tSZ}).
%Thus, our work represent the state-of-art constraints using cross-correlation between CMB lensing and DLAs/QSOs.}
Moreover, we provide a step-by-step procedure of the cross-correlation
in details for the future references.

Our results have also shown a smaller QSO bias comparing to the auto-correlation results.
We have not yet found a convincing explanation for this discrepancy
beyond statistical fluctuations.
The systematics induced by the \textit{Planck} and the lensing reconstruction may account for the fluctuation. Systematics related to incompleteness, random catalog generation, masking and so on may also contribute to this difference, which should be carefully considered when using the auto-correlation approach \citep[e.g.][]{Reid_2015,Geach_2019}.
However, the bias measurement of DLAs remains effectively immune to the unknown fluctuation of quasar bias, since $\bdla$ is subtracted from the combined lensing signal,
i.e. a combination of weak lensing effect caused by both quasars and its foreground DLAs, and these quasars would share the same systematics as the selected BOSS quasars.

In this paper, we propose a meaningful upper limit for the DLA bias and our results favours a DLA halo mass of $M_{\rm{halo}}<10^{\upmdla} M_\odot h^{-1}$.
Some observations indicate that metal-rich DLAs are
hosted by massive halos with mass reaching $10^{12}\ M_\odot$ \cite[e.g.,][]{Neeleman2018},
while a number of hydrodynamical simulations and astrophysical models \citep[e.g.][]{DLA_sim_1} prefer that DLAs generally reside in smaller halos. Resolving the discrepancy in the clustering strength of DLAs is important to determine the nature of DLAs. %and helps for further novel applications.
\cite{Lyman_forest_2012} measured $\bdla$ by the cross-correlation of DLA and the Ly$\alpha$ forest. The value of the mean bias they measured for DLAs is $\bdla=2.17\pm0.20$, indicating a halo mass of $\sim 10^{12}M_\odot$. Using the same technique, \citet{Lyman_forest_2018} updated the result using the BOSS Data Release 12. Comparing with \cite{Lyman_forest_2012}, \citet{Lyman_forest_2018} favor a lower bias, yielding $\bdla = 2.00\pm0.19$ and corresponding to a halo mass of $\sim 4\times 10^{11} h^{-1}M_\odot$.
Using the CMB lensing with \textit{Planck} 2018 convergence map, although the error bar of our measurement is much larger than the DLA-Ly$\alpha$ forest cross correlation results, our measurement provides an independent and a good constraint on
the upper limit. The upper limit we pose is that the DLA halo mass is lower than 10$^{\upmdla}\ M_\odot h^{-1}$ at a 90\% confidence level, and this is consistent with 
%1$\sigma$ the upper limit of 
the Ly$\alpha$ cross-correlation measurement. The relatively low resolution (5-10 arcmin), relatively low signal-to-noise of \textit{Planck} CMB lensing map, and the current limited sample sizes of selected quasars and DLAs may account for the large uncertainty. We expect new surveys such as DESI\footnote{\url{https://www.desi.lbl.gov/}} to provide a larger quasar and DLA database, and improvements in CMB lensing, such as ACT\footnote{\url{https://act.princeton.edu/}}, Simons Observatory (SO)\footnote{https://simonsobservatory.org/} and CMB-S4\footnote{\url{https://cmb-s4.org/}}, will also help a lot for a more precise bias measurement.

\section{Conclusion}\label{sec:conclusion}
We measure the bias of DLAs by studying the cross-correlation between the selected quasars with DLAs in their sightline and CMB lensing. We perform the measurement based on the bias measured for a BOSS subset. We find a quasar bias $\bqso=\BOSSbqso$ for the BOSS subset, corresponding to a quasar halo mass of log$(M/M_\odot h^{-1})=\BOSSmqso$ at a median redshift of $z_{\rm m}\approx2.51$. Our measurement of quasar bias is 2-$\sigma$ smaller than the fiducial bias model \citep{BOSS_auto}.  Correlated systematics between the CMB lensing map and the quasar sample might account for such a low $\bqso$, but we have not yet found an explanation convincing enough. Despite its large uncertainty, our result provides a good constraint on the upper limit of DLAs halo mass. We find $\bdla \leq \upbdla$ and log$(M/M_\odot h^{-1})\leq \upmdla$ at a confidence level of 90\%, consistent with previous work \citep{Lyman_forest_2012,Lyman_forest_2018}.

A simple null test is performed for the BOSS subset and the selected quasars with DLAs, by cross-correlating the CMB lensing convergence map on one part of the sky and the quasar map on another part of the sky. For the robustness of our measurement, we also repeat our analysis using a redshift-dependent bias model as the fiducial, and on a CMB lensing map with thermal-SZ signal deprojected (see Appendix).
%\sout{ CMB lensing field is a good tracer for the large-scale distribution of matter in the universe, especially objects at high redshift ranges. This cross-correlation approach is expected to be more precise and less prone to statistical uncertainties if samples of large size are used.}
We expect future survey and CMB detection will improve the precision and accuracy of the DLA bias measurement..

\section{acknowledgement}
We acknowledge the referee of this paper, David Alonso, for the thoughtful and insightful comments which greatly improved the paper. ZC and XL are supported by the National Key R\&D Program of China (grant No. 2018YFA0404503). We thank Jiashu Han, Dandan Xu, John Andrew Peacock and Andreu Font-Ribera for helpful discussions. We also acknowledge the support of time and computing resources from DoA at Tsinghua University, and the kind reply of PLA Helpdesk\footnote{\url{https://support.cosmos.esa.int/pla/}}. Our work is based on observations made by the Planck satellite and the Apache Point Observatory. The Planck project\footnote{\url{http://www.esa.int/Planck}} is funded by the member states of ESA and NASA. The SDSS-\uppercase\expandafter{\romannumeral3}\footnote{\url{http://www.sdss.org}} project is funded by the participating institutions, the National Science Foundation, the United States Department of Energy and the Alfred P. Sloan Foundation.

\software
{
Healpy \citep{healpix}, 
COLOSSUS \citep{colossus}, 
Scipy \citep{scipy}, 
CAMB \citep{Lewis_2000},
emcee \citep{emcee}.
}

\appendix

\section{Test for bias evolution model}\label{Laurent17}
 Adopting a model for the bias evolution with redshift is a good test for the robustness of our result. While \citet{BOSS_auto} found a roughly redshift-independent bias for the BOSS sample, other works \citep{Shen,Laurent2017} found that the quasar bias sharply evolves with redshift.  If the bias is redshift-dependent, the small mismatch in redshift distribution between the BOSS subset and the DLA quasars (Fig.\ref{fig:distribution}) could create a difference in their clustering, spuriously mimicking a change in the DLA bias. We test this possibility here by adopting an evolving bias model as an alternative to Eq.\ref{scaling_a}. 
 \cite{Laurent2017} measured the quasar correlation function of the first year of the eBOSS quasar sample, and provided a bias-redshift model by combining their results and the bias measured with the BOSS sample:

\begin{equation}\left.\bqso(z)=\alpha\left[(1+z)^{2}-6.565\right)\right]+\beta\end{equation}

\noindent with $\alpha=0.278\pm0.018,\beta = 2.393 \pm 0.042$. Despite the different sample selection, we adopt the central values of these parameters and use this bias model as the fiducial quasar bias in our analysis, and assume DLA bias remains a constant over the selected redshift range.

With this bias evolution model, we get $a=0.53\pm0.16$, corresponding to $\bqso=a\cdot b_{\rm{fid}}(z_{\rm m})=2.13\pm0.63$ and a characteristic halo mass of log$(M/M_\odot h^{-1})=11.45_{-0.81}^{+0.48}$ at a median redshift of $z_{\rm m}\approx2.51$. Furthermore, it gives $\bdla=1.76^{+1.40}_{-1.11}$. The 90\% upper limit of DLA bias is $3.59$, corresponding to a halo of log$(M/M_\odot h^{-1})=12.50$.

This result is consistent with our previous analysis.

\section{Test for tSZ-free lensing signal}\label{non-tSZ}

Thermal Sunyaev-Zel'dovich (tSZ) contamination is considered to be a potential systematic to CMB lensing cross-correlation analyses \citep{van2014,Osborne2014}. The Sunyaev-Zel'dovich (SZ) effect is caused by the inverse Compton scattering of CMB photons off hot electrons in the deep potential wells of galaxy clusters. Although the tSZ signal mainly comes from low-redshift structure, and is not expected to be a large source of contamination in our high-redshift analysis, it is worth testing the robustness of our result by repeating the bias measurement with tSZ-free lensing map. The Planck Collaboration released a lensing reconstruction map on \texttt{SMICA} foreground-cleaned maps, where lensing signal is estimated from temperature (TT) only with tSZ signal deprojected.

\begin{figure*}[htbp]
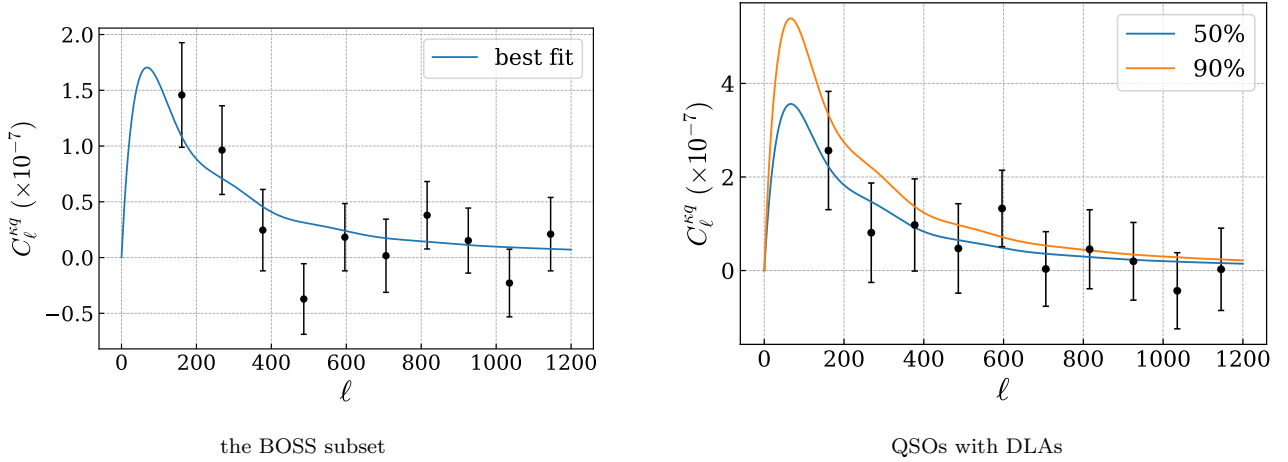

\gridline{
        \fig{./tSZ_test/DR12_spectrum_non-tSZ.pdf}{0.45\textwidth}{the BOSS subset}
        \fig{./tSZ_test/DLA_spectrum_non-tSZ.pdf}{0.45\textwidth}{QSOs with DLAs}
          }
\caption{Cross-power spectrum on tSZ-free lensing signal with the BOSS subset (the left panel) and with the QSOs with DLAs (the right panel) over the redshift range of $2.2<z<3.4$. In the left panel, the solid line is the best-fit theoretical curve. In the right panel, the two solid lines denote the theoretical cross-power spectra at 50\% and 90\% upper limits of $\bdla$.
\label{fig:non-tSZ spectrum}}
\end{figure*}

In the absence of tSZ bias, we get $a=0.68\pm0.19$, $\bqso=a\cdot b_{\rm fid} = 2.45\pm0.70, log(M/M_\odot h^{-1})=11.72_{-0.68}^{+0.42}$ for the BOSS quasars.
%, and $\bdla=1.28_{-0.69}^{+0.76}$. log$(M/M_\odot h^{-1})=10.42_{-16.06}^{+1.14}$ for DLA sample.
The 90\% upper limit of DLA bias is $3.78$ and that of DLA halo mass is log$(M/M_\odot h^{-1})=12.58$. Fig.\ref{fig:non-tSZ spectrum} shows the data and best-fit spectra.

The consistency shows that foreground contamination is subdominant. Since the tSZ-free analysis is based on temperature maps only and lacking polarization information, we take the measurement on the MV estimated lensing signal as our primary conclusion.

\begin{figure*}[htbp]
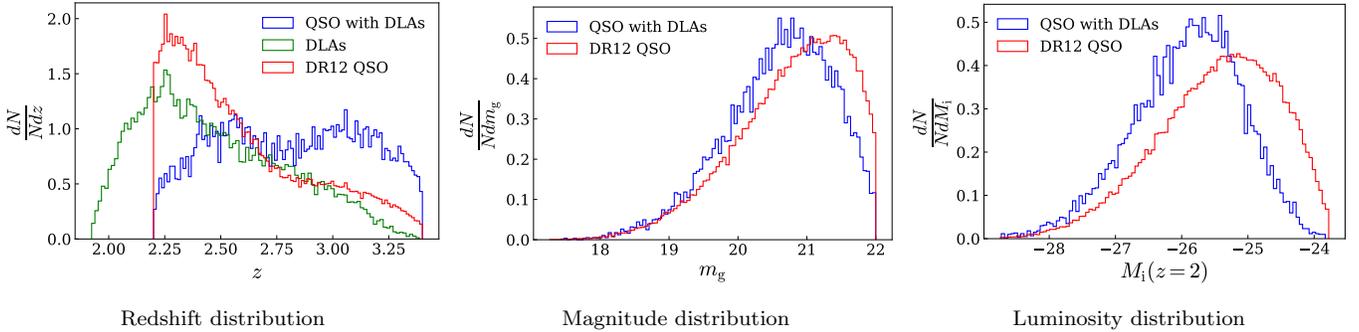

\gridline{
        \fig{./Mingfeng/Redshift_Mingfeng.pdf}{0.33\textwidth}{Redshift distribution}
        \fig{./Mingfeng/Magnitude_Mingfeng.pdf}{0.33\textwidth}{Magnitude distribution}
        \fig{./Mingfeng/Luminosity_Mingfeng.pdf}{0.33\textwidth}{Luminosity distribution}
          }
\caption{The normalised redshift, magnitude and luminosity distributions of the selected BOSS quasar and DLA samples in the \citet{Mingfeng_2020} catalog.
\label{fig:distribution_Mingfeng}}
\end{figure*}

{
\section{Test for another DLA catalog}\label{Mingfeng}

\citet{Mingfeng_2020} provided an revised DLA catalog using Gaussian processes. Their pipeline improved the ability to detect multiple DLAs along a single sightline. They analysed 158,825 Lyman-$\alpha$ spectra selected from SDSS DR12 and present updated estimates for the statistical properties of DLAs, including
the column density distribution function, line density, and neutral
hydrogen density. We apply our sample selection criteria to their DLA catalog as described in \S \ref{sample_selection} except the DLA \texttt{confidence} (they don't have such a parameter in their catalog), and get 12,368 quasars with 17,836 foreground DLAs. Fig. \ref{fig:distribution_Mingfeng} shows the properties of the selected sample. Also we repeat the steps in \S \ref{cov} to calculate the corresponding covariance matrix.

With this new DLA catalog we get $a=0.78\pm0.20$, $b_{\rm QSO}=2.79\pm0.72$ for the BOSS quasars. As for DLAs, we obtain $b_{\rm DLA}=2.71^{+1.28}_{-1.25}$. This result is perfectly consistent with that in \citet{JCAP_DLA}. The 90\% upper limit of $b_{\rm DLA}$ is 4.34, corresponding to a halo of log$(M/M_\odot h^{-1})=12.69$. Fig.\ref{fig:Mingfeng spectrum} shows the data and best-fit spectra.
}

\begin{figure*}[htbp]
\gridline{
        \fig{./Mingfeng/DR12_spectrum_Mingfeng.pdf}{0.45\textwidth}{the BOSS subset}
        \fig{./Mingfeng/DLA_spectrum_Mingfeng.pdf}{0.45\textwidth}{QSOs with DLAs}
          }
\caption{Cross-power spectrum on CMB lensing signal with the BOSS subset (the left panel) and with the QSOs with DLAs in the \citet{Mingfeng_2020} catalog (the right panel) over the redshift range of $2.2<z<3.4$. In the left panel, the solid line is the best-fit theoretical curve. In the right panel, the two solid lines denote the theoretical cross-power spectra at 50\% and 90\% upper limits of $\bdla$.
\label{fig:Mingfeng spectrum}}
\end{figure*}

\section{Apodization}\label{apod}
We use mocks to to decide a proper kernel size.
    For each mock, we start with theoretical %\yl{more details?}
    curves $C_\ell^{\kappa \kappa}$, $C_\ell^{qq}$ and $C_\ell^{\kappa q}$ given by:
    \begin{equation}
    \begin{array}{l}
    C_\ell^{\kappa\kappa}=\int \frac{dz}{c}\frac{H(z)}{\chi^2(z)}W^2(z)P_{mm}(k=\frac{\ell}{\chi(z)},z),\\
    C_\ell^{qq}=\int \frac{dz}{c}\frac{H(z)}{\chi^2(z)}f^2(z)P_{mm}(k=\frac{\ell}{\chi(z)},z),\\
    C_\ell^{\kappa q}=\int \frac{dz}{c}\frac{H(z)}{\chi^2(z)}W(z)f(z)P_{mm}(k=\frac{\ell}{\chi(z)},z).
     \end{array}
    \end{equation}
%\yl{Let's move the two auto-power equations to Sec3.4 and leave a
%pointer to that here.}
 See more details of these equations in \S \ref{sub:theory}.
    We draw a sample of $2\times10^8$ quasars with the bias $b_{\rm QSO}=2.5$
%    \yl{need to unify symbols on quasar bias}
    uniformly distributed over
    the redshift range of $2.2\leq z \leq3.4$.
    We then generate Gaussian-distributed correlated CMB
    lensing and quasar fields through \texttt{healpy.synfast} following
    Eq.13 and Eq.14 in \cite{mask_synfast}:
    \begin{equation}\begin{array}{l}
    a_{\ell m}^{\mathrm{\kappa \kappa}}=\xi_{a}\left(C_{\ell}^{\mathrm{\kappa \kappa}}\right)^{1 / 2} \\
    a_{\ell m}^{qq}=\xi_{a} C_{\ell}^{\mathrm{\kappa q}} /\left(C_{\ell}^{\mathrm{\kappa \kappa}}\right)^{1 / 2}+\xi_{b}\left(C_{\ell}^{qq}-\left(C_{\ell}^{\mathrm{\kappa q}}\right)^{2} / C_{\ell}^{\mathrm{\kappa \kappa}}\right)^{1 / 2}
    \end{array}\end{equation}
    where $\xi$ denotes a random amplitude so that $\left\langle\xi \xi^*\right\rangle=1$ and $\left\langle\xi\right\rangle=0$, yielding

    \begin{equation}\begin{aligned}
    \left\langle a_{\ell m}^{\mathrm{\kappa \kappa}} a_{\ell m}^{\mathrm{\kappa \kappa} *}\right\rangle &=C_{\ell}^{\mathrm{\kappa \kappa}}, \\
    \left\langle a_{\ell m}^{\mathrm{\kappa \kappa}} a_{\ell m}^{qq *}\right\rangle &=C_{\ell}^{\mathrm{\kappa q}}, \\
    \left\langle a_{\ell m}^{qq} a_{\ell m}^{qq *}\right\rangle &=C_{\ell}^{qq}.
    \end{aligned}\end{equation}

    We multiply the synthetic Gaussian maps by the original binary
    masks, to ensure that pixels without data remain empty, such as
    regions around the Galactic plane.
    We then apply the apodized mocks with FWHM given in Table
    \ref{tab:kernel_size} before measurement of cross-power spectra.
    We repeat these steps on 200 Gaussian mocks, average them together
    and bin them into 10 bands.
    By comparing the binned average reconstructed spectra
    ${\hat{C}}_i^{\kappa q}$ and the binned theoretical template
    $C_i^{\kappa q}$ and calculating the mean squared error (MSE) of the residuals:
%    \yl{strictly speaking this is not a variance. So maybe rename them to MSE here and in the table}
    \begin{equation}
        {\rm{MSE}}=\sum_{i=1}^{N}\frac{(\hat{C}_i^{\kappa q} - C_i^{\kappa q})^2}{N}
    \end{equation}
    where $N=10$.

\bibliography{cross-correlation}{}
\bibliographystyle{aasjournal}



\end{document}